\pdfoutput=1

\documentclass[11pt]{article}

\usepackage[final]{acl}

\usepackage{times}
\usepackage{latexsym}
\usepackage{hyperref}

\usepackage[T1]{fontenc}

\usepackage[utf8]{inputenc}

\usepackage{microtype}
\usepackage{amsmath}
\usepackage{pbox}
\usepackage{chngpage}
\usepackage{minitoc}
\usepackage{longtable}
\usepackage{comment}
\usepackage{tcolorbox}
\usepackage{booktabs}
\usepackage{subcaption}
\usepackage{amssymb}
\usepackage{multirow}
\usepackage[linesnumbered,ruled,vlined]{algorithm2e}

\usepackage{inconsolata}

\usepackage{graphicx}

\title{SkewRoute: Training-Free LLM Routing for Knowledge Graph Retrieval- Augmented Generation via Score Skewness of Retrieved Context}
\author{
	  Hairu Wang$^{1,3}$\thanks{Equal contribution.}, Yuan Feng$^{1,3}$\footnotemark[1], Yukun Cao$^{1,3}$,
      Xike Xie$^{2,3}$\thanks{Corresponding author.}, S Kevin Zhou$^{2,3}$ \\
	  $^1$School of Computer Science, University of Science and Technology of China, China\\ $^2$School of Biomedical Engineering, USTC, China\\ $^3$Suzhou Institute for Advanced Reasearch, USTC, China\\
	  \footnotesize\texttt{\{hrwang00,yfung\}@mail.ustc.edu.cn}, 
	  \footnotesize\texttt{\{xkxie,skevinzhou\}@ustc.edu.cn}
}

\begin{document}
\maketitle
\begin{abstract}
Large language models excel at many tasks but often incur high inference costs during deployment. To mitigate hallucination, many systems use a knowledge graph to enhance retrieval-augmented generation (KG-RAG). However, the large amount of retrieved knowledge contexts increase these inference costs further. A promising solution to balance performance and cost is LLM routing, which directs simple queries to smaller LLMs and complex ones to larger LLMs.
However, no dedicated routing methods currently exist for RAG, and existing training-based routers face challenges scaling to this domain due to the need for extensive training data. We observe that the score distributions produced by the retrieval scorer strongly correlate with query difficulty. 
Based on this, we propose an extremely simple yet effective routing framework, the first specifically designed for KG-RAG that efficiently balances performance and cost in a plug-and-play manner. It delivers over 3x higher routing effectiveness while reducing runtime to less than 0.001x compared to existing methods. 
Our code is available at \url{https://github.com/hrwang00/SkewRoute}.
\end{abstract}

\section{Introduction}
\label{sec:intro}
Large Language Models (LLMs) \cite{openai2024gpt4technicalreport, llama3.1, qwen2025, geminiteam2024geminifamilyhighlycapable} exhibit remarkable capabilities across a wide range of tasks. However, LLMs often hallucinate due to a lack of up-to-date or domain-specific knowledge \cite{hong2023faithful, wang2023knowledge}. To mitigate this, LLMs are often augmented with retrieved external knowledge \cite{asai2024selfrag, yu2024rankrag, gao2024modularragtransformingrag} during application. 
Notably, many methods use externally structured Knowledge Graphs (KGs) as external bases, retrieving query-relevant context to enhance subsequent LLM generation \cite{luo2024reasoning,he2024gretriever,hu2024graggraphretrievalaugmentedgeneration,li2025simple}. KGs typically represent knowledge as triples. In Retrieval-Augmented Generation (RAG), each triple serves as a knowledge context unit, scored based on its semantic and structural relevance to the given query. Subsequently, a set of high-scoring candidates is selected and concatenated to provide the LLM with supplementary knowledge, thereby enhancing generation quality.
For instance, SubgraphRAG \cite{li2025simple} employs a lightweight MLP to score independent triple as knowledge context and achieves state-of-the-art performance by concatenating these contexts.

\begin{figure}[t]
    \centering
    \includegraphics[width=0.99\linewidth]{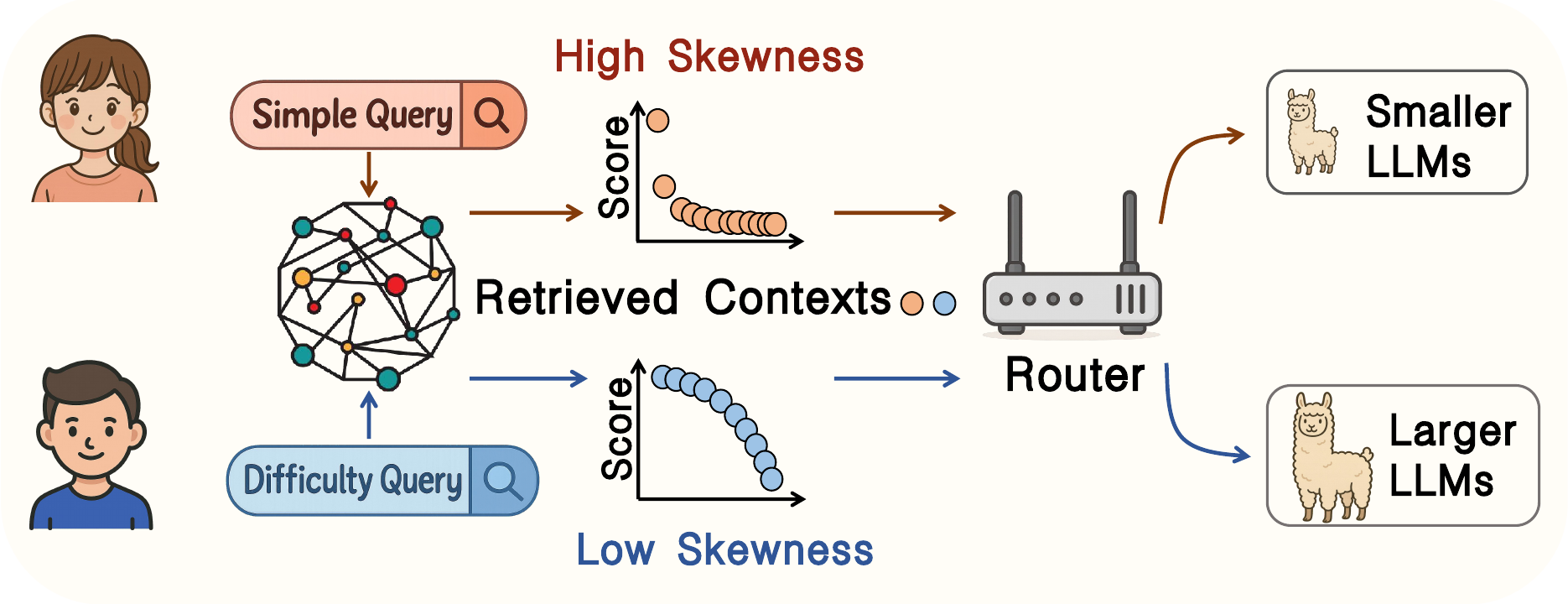} 
	\caption{\textbf{A Training-Free Routing Framework for LLMs in KG-RAG.} Scores of retrieved contexts sorted in descending exhibit distinct skewness pattern. The framework utilizes the score skewness of retrieved contexts to route requiring no training.}
    \label{fig:framework}
\end{figure}

Given the autoregressive nature of LLMs, inference costs are directly proportional to the number of tokens \cite{hao2025omnikv, snell2025scaling}. While KG-RAG mitigates hallucinations and improves generation quality, it dramatically expands input tokens due to the increased retrieved context. As shown in Figure \ref{fig:statistic_data:tokens}, answering a question directly with an LLM requires 62 tokens per question. 
However, retrieving 100 triples using SubgraphRAG increases this to 1873 tokens per question, resulting in an over 30x increase in input token consumption. This significant overhead hinders practical LLM application. 

To reduce LLM inference costs in applications, recent studies have focused on LLM routing \cite{aggarwal2024automix, ding2024hybrid, ong2025routellm, feng2025graphrouter}, routing simple questions to smaller LLMs and complex questions to larger LLMs. This is primarily because the performance benefits of increased model scale exhibit diminishing returns; 
for example, Qwen2.5-14B-Instruct incurs less than twice the inference cost of Qwen2.5-7b-Instruct while delivering a 7.45\% performance increase. In contrast, Qwen2.5-72b-Instruct requires almost 7x the inference cost of Qwen2.5-14b but only achieves a 2.12\% improvement, as illustrated in Figure \ref{fig:statistic_data:price}.
However, while recent routing methods have been explored in direct LLM QA applications, their application and extensibility to RAG domains is limited, primarily due to two reasons:

\begin{figure}[t] 
    \centering 
    \begin{subfigure}[b]{0.49\linewidth} 
        \centering 
        \includegraphics[width=\linewidth]{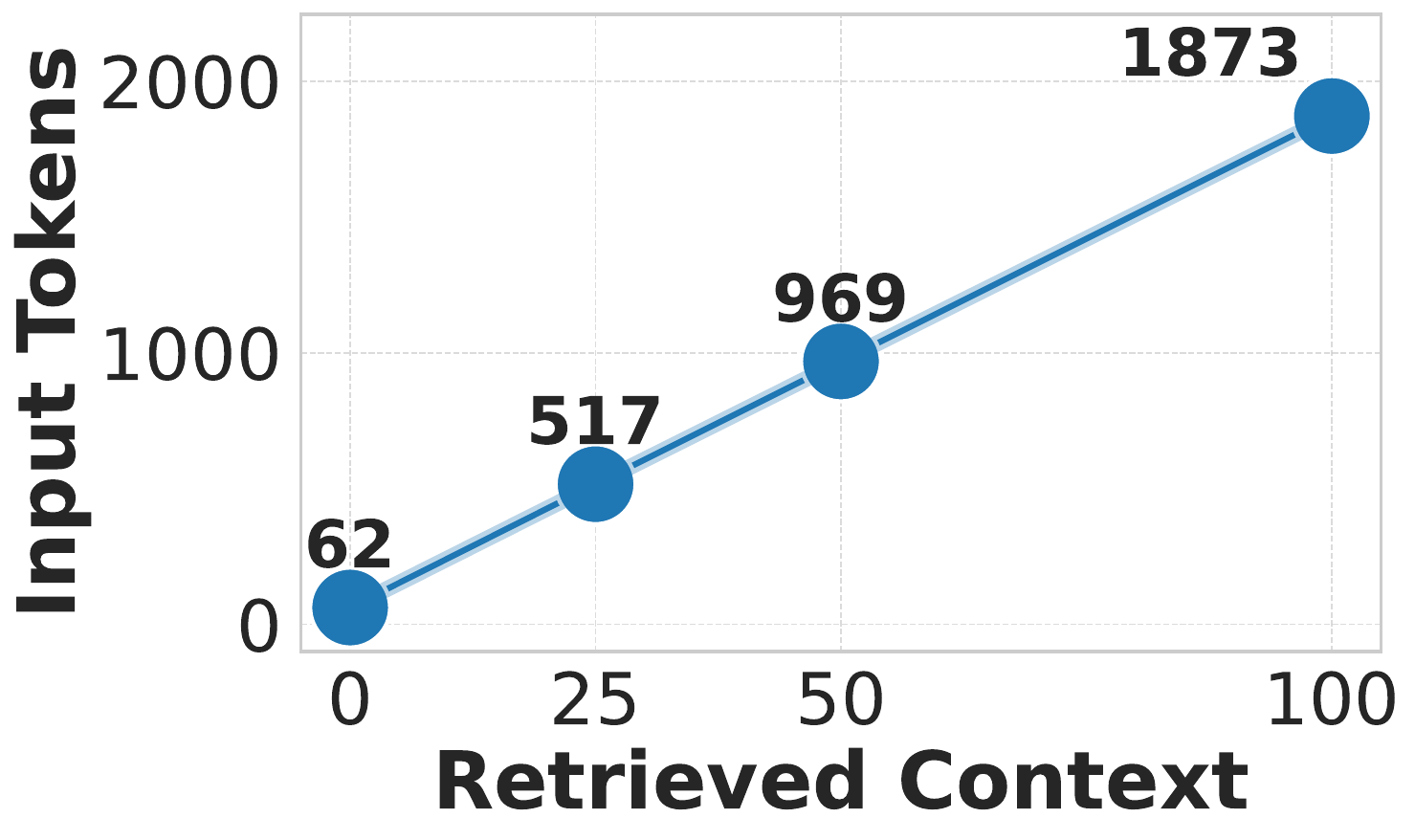}
        \caption{Token vs. Context}
        \label{fig:statistic_data:tokens}
    \end{subfigure}
    \hfill 
    \begin{subfigure}[b]{0.49\linewidth} 
        \centering 
        \includegraphics[width=\linewidth]{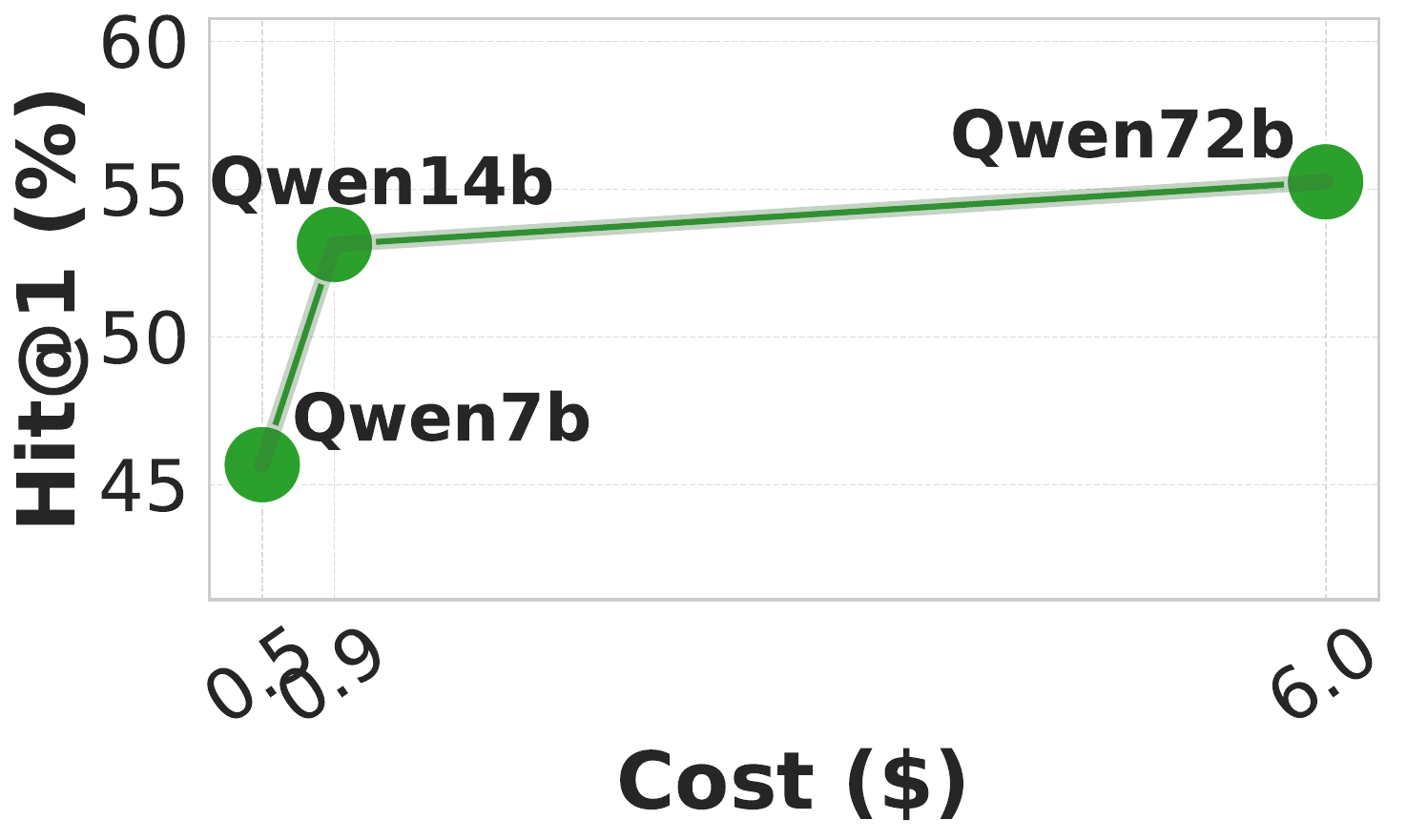}
        \caption{Performance vs. Cost}
        \label{fig:statistic_data:price}
    \end{subfigure}
    \caption{\textbf{Token and Performance-Cost Statistics on CWQ.} (a) illustrates how input tokens varies with retrieved contexts in KG-RAG. (b) presents inference cost and performance on LLM cloud service platform of different LLM scales.}
    \label{fig:statistic_data} 
\end{figure}

\textbf{(1) Training-Based Routing is Difficult to Construct:}
Existing methods rely on training a classifier to route between multiple LLMs, which inherently requires substantial amounts of high-quality labeled data specifying the optimal model for a given query. Acquiring this data via expert annotation is costly and doesn't generalize well to new datasets, typically necessitating retraining or fine-tuning.

\textbf{(2) Knowledge Source Shift Drives Routing Objective Transformation:}
In direct LLM QA, the LLM router relies on the internal knowledge differences between models for routing. This differs fundamentally from common RAG scenarios, where knowledge dependence is shifted to externally retrieved knowledge. This reduces the importance of inherent model knowledge and increases focus on the LLMs' reasoning ability over external knowledge. This difference in routing objectives makes existing routing methods unsuitable for RAG.

These challenges highlight the significant hurdles of effective routing LLMs in RAG scenarios, which raises a critical question: \textit{Can we develop a training-free routing method, specifically tailored for RAG, that routes by focusing target reasoning on external knowledge sources?}

By analyzing the score distribution of retrieved contexts in KG-RAG, we surprisingly find significant differences for different queries, as shown in Figure \ref{fig:framework}. The score distribution for some queries exhibits high skewness, characterized by power-law behavior,which is widely observed across many fields. \footnote{For a detailed introduce of power-law, please see Section~\ref{sec:pre}.} \cite{takahashi-tanaka-ishii-2019-evaluating, neumann2024alphazeroneuralscalingzipfs, 10499867, feng2024mayfly}. 
A handful of contexts dominate with high scores, leaving a long tail of many lower-scoring ones.
This indicates that only a few knowledge contexts are highly relevant to the query, while the majority are not. However, other queries show a distinctly different, low skewness distribution, where a substantial portion of the contexts exhibit high relevance scores. This skewness variation is strongly correlated with the difficulty of reasoning over external knowledge. For instance, simple questions may require only simple matching within a single context, while difficult questions often demand multi-hop reasoning across multiple knowledge contexts. Thus, the varying skewness of the score distribution may serve as an indicator of query difficulty.

Inspired by the score skewness of retrieved information in KG-RAG, we first investigate the correlation between skewness patterns and query difficulty. Building on this insight, we propose the first training-free routing framework to address the existing routing challenges in KG-RAG scenarios. This framework employs a lightweight and efficient workflow, as shown in Figure \ref{fig:framework}, that leverages the skewness pattern of the retrieved context scores to dynamically route queries. Simpler queries are assigned to smaller LLMs, while more difficult queries are routed to larger LLMs, effectively balancing performance and cost. Our main contributions are summarized as follows:

\begin{list}{$\bullet$}
	{   \setlength{\itemsep}{0pt}
		\setlength{\parsep}{3pt}
		\setlength{\topsep}{3pt}
		\setlength{\partopsep}{0pt}
		\setlength{\leftmargin}{1.5em}
		\setlength{\labelwidth}{1em}
		\setlength{\labelsep}{0.5em} }
	\item To the best of our knowledge, this work is the first to introduce a training-free routing framework tailored for KG-RAG, leveraging the skewness patterns in retrieved scores to dynamically assess query difficulty, offering a novel perspective for future efficient RAG research
	\item We propose a novel training-free framework that fully leverages the skewness pattern of retrieved scores from external knowledge bases, enabling it to balance cost and performance between different LLMs with high flexibility.
	\item Extensive  evaluations  across multiple LLM scales demonstrate the superiority and efficiency of our proposed routing framework, highlighting its strong generalization capability and practical value for cost-effective LLM deployment.
\end{list}

\section{Preliminaries}
\label{sec:pre}
\textbf{Power-law.}
Power-law \cite{song2024powerinferfastlargelanguage, havrilla2024understandingscalinglawsstatistical}, ubiquitous in nature, reveals an inverse relationship between scores and their descending ranks: $S(n) \;=\; \frac{C}{n^{\alpha}}$.
In this formulation, $n$ denotes the rank and $S(n)$ is the corresponding score. 
In log-log space, $logS(n)$ against $log(n)$ approximates a straight line with slope $-\alpha$. 
While in linear space, it appears as a few high scores followed by a long tail of lower ones. 
In KG-RAG, we observe that for certain queries, the scores of retrieved contexts also follow this power-law behavior. 

\noindent\textbf{KG-RAG.}
RAG integrates contexts retrieved from external knowledge bases to enhance LLMs.
In KG-RAG, the knowledge bases are KGs, and the retrieved contexts take the form of triples $\{\tau = (h, r, t)\}$.
Existing research on KG-RAG generally involves two stages: retrieval and generation.
For example, the state-of-the-art KG-RAG method, SubgraphRAG trains a scorer $\mathcal{R}$ to select a subset of triples $T = \{\tau_1, \tau_2, \dots, \tau_K\}$ by their scores \(W = \{s_1, s_2, \dots, s_K\}\) relevant to $q$ from $G$ in the retrieval phase. 
And then feeds $q$, an instruction $I$, and $T$ to a generator $\mathcal{G}$ (typically an LLM) to produce the final answer: $\mathcal{G}(I, T, q) \rightarrow a$, where $a \in \mathcal{A}$.
Knowledge Graph Question Answering (KGQA) is often used to evaluate KG-RAG, aiming to obtain answers from KGs given query $q$, where $q \in Q$. 

\noindent\textbf{LLM Route.}
Based on scale, LLMs are broadly categorized as larger LLMs and smaller ones.
Given a query $q \in Q$, a larger LLM is formally represented as a function $F_L(I, T, q)=a$, mapping the query $q$ and retrieved information $T$ to an answer $a$, while a smaller LLM is denoted as $F_S(I, T, q)=a$.
Larger LLMs deliver high quality with significant cost. In contrast, smaller LLMs are less capable but substantially reduce overhead.
For queries that smaller LLMs fail but larger LLMs handle correctly, using the larger LLMs boosts performance. However, for queries both LLMs answer correctly, relying on the larger LLMs incurs a substantial extra cost.
Unlike previous works \cite{aggarwal2024automix, ding2024hybrid, ong2025routellm} that train a binary classifier to route between smaller LLMs and larger LLMs,
we design training-free router $\mathcal{L}$ on the scores $W$ of retrieved contexts for given query $q$, $\mathcal{L}:\mathcal{W}\rightarrow \{F_S, F_L\}$, where $\mathcal{W}$ is the space of possible $W$ and $\theta$ is a threshold.
\[
a = 
\begin{cases} 
    F_S(I, T, q) & \text{if } \mathcal{L}(W, \theta) \le 0 \\ 
    F_L(I, T, q) & \text{if } \mathcal{L}(W, \theta) > 0
\end{cases}
\]

\section{Methodology}
\subsection{Skewness of Retrieved Context Scores}
\begin{figure}[htbp] 
    \centering 
    \begin{subfigure}[b]{0.49\linewidth} 
        \centering 
        \includegraphics[width=\linewidth]{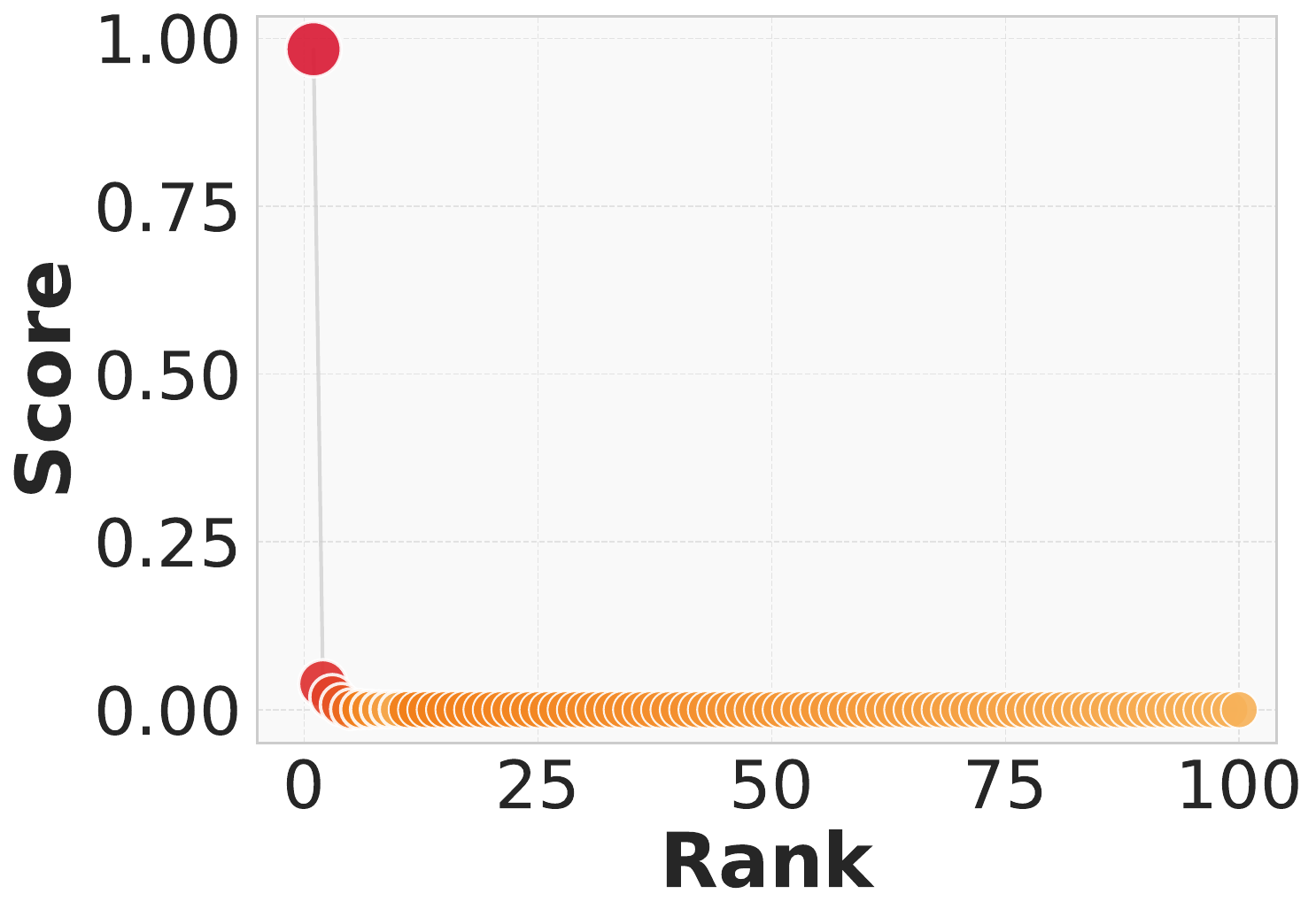}
        \caption{High Skewness} 
        \label{fig:analysisonV:highskew:origin} 
    \end{subfigure}
    \hfill 
    \begin{subfigure}[b]{0.49\linewidth}
        \centering
        \includegraphics[width=\linewidth]{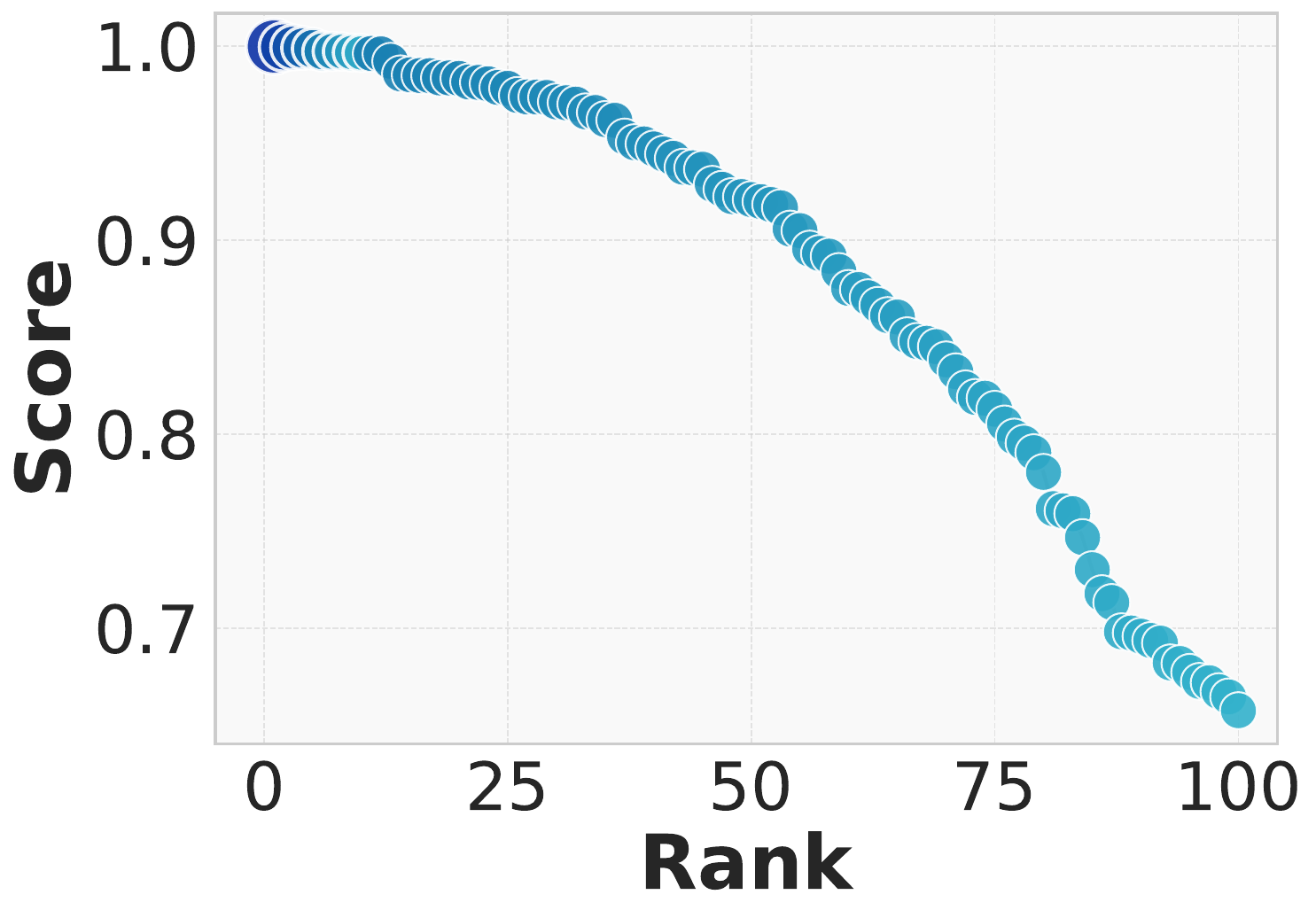}
        \caption{Low Skewness}
        \label{fig:analysisonV:lowskew:origin}
    \end{subfigure}
    \begin{subfigure}[b]{0.49\linewidth}
        \centering
        \includegraphics[width=\linewidth]{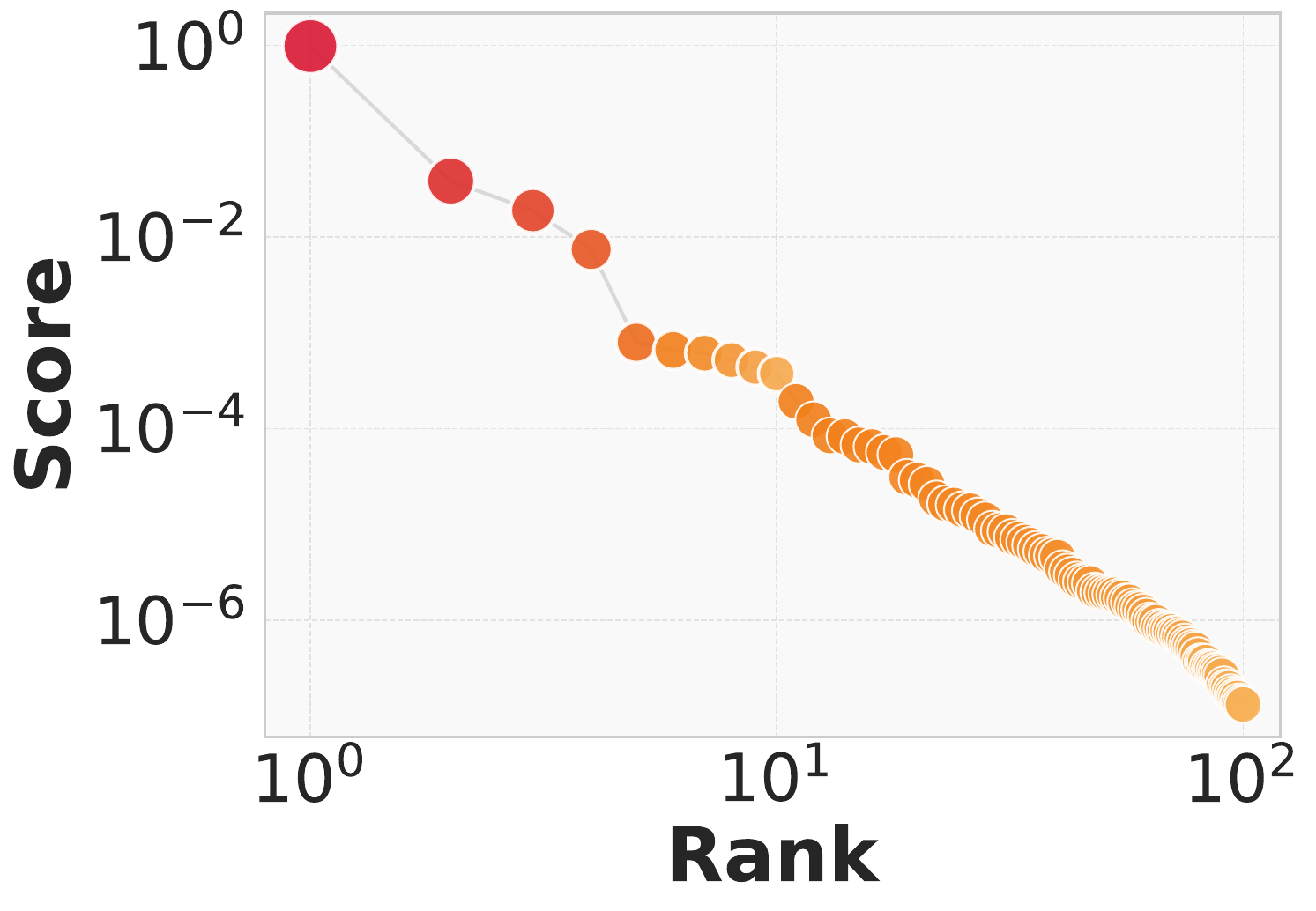}
        \caption{High Skewness}
        \label{fig:analysisonV:highskew}
    \end{subfigure}
    \hfill 
    \begin{subfigure}[b]{0.49\linewidth}
        \centering
        \includegraphics[width=\linewidth]{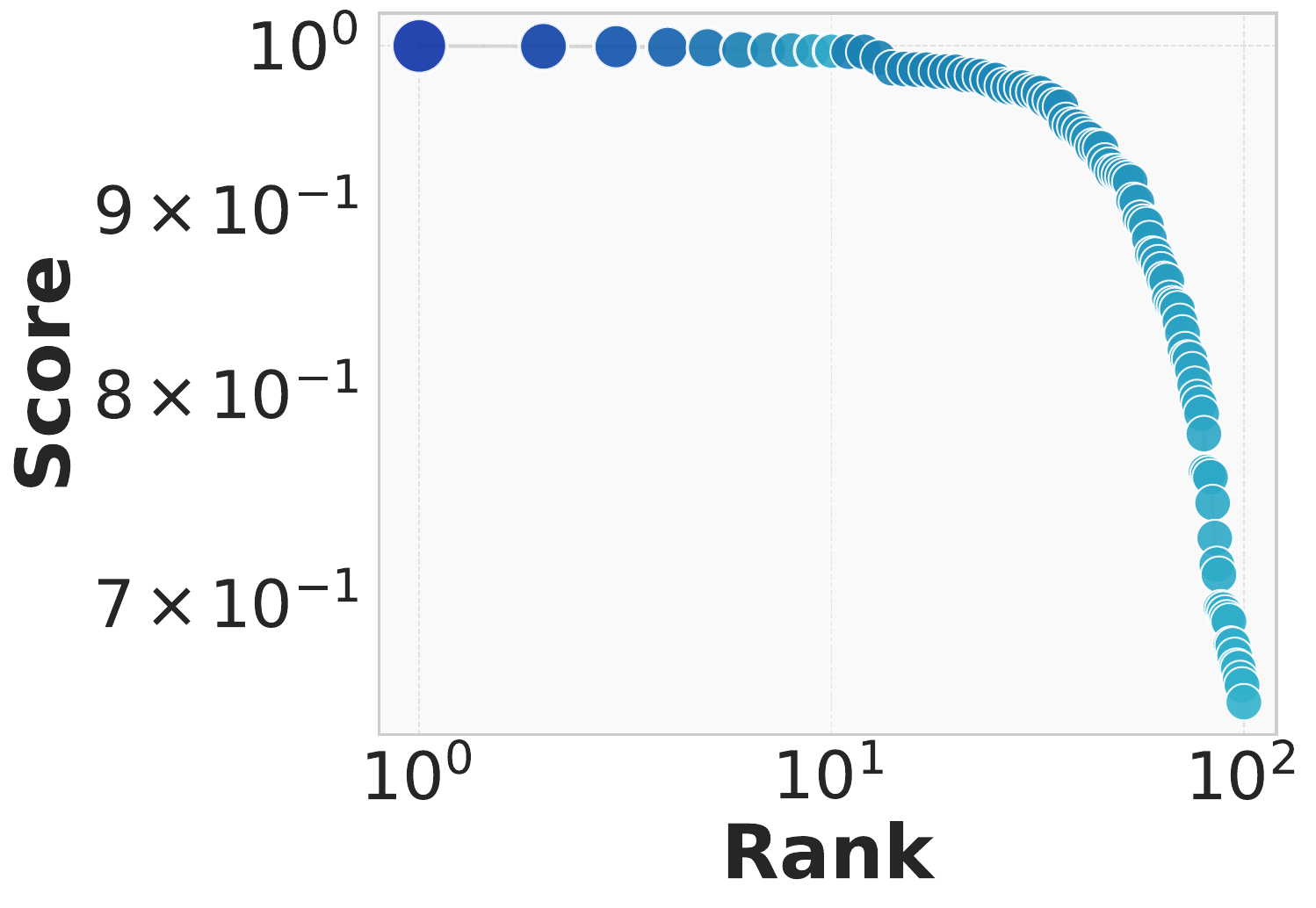}
        \caption{Low Skewness}
        \label{fig:analysisonV:lowskew}
    \end{subfigure}
    \caption{\textbf{Score of Retrieved Contexts in CWQ.} (a)(b) are plotted in linear coordinates, while (c)(d) employ a log-log scale. }
    \label{fig:analysisonV} 
\end{figure}

The KG-RAG method typically trains a scorer that assigns higher scores to contexts relevant to the query and lower scores to irrelevant ones. Based on these scores, the top-$K$ contexts are selected to enhance LLM generation, making the quality of this selection directly affect the final answer quality. In reality, queries can be categorized as simple or difficult \cite{berant-etal-2013-semantic,yih-etal-2014-semantic,complex2simple,EmbedKGQA}. Simple queries often require only single-hop information matching within one knowledge context to find the answer, while difficult queries usually need multi-hop reasoning across multiple knowledge contexts. To explore this, we visualize the score distributions assigned by the scorer across different queries and find significant differences that strongly correlate with the query difficulty.

Figure \ref{fig:analysisonV} presents two typical examples of the top-100 context score distributions for different queries, showing clear and significant differences. 
Power-law distribution is highly skewed with long-tail.
In Figure \ref{fig:analysisonV:highskew:origin} and Figure \ref{fig:analysisonV:highskew}, the distribution exhibits power-law behavior and approximates a straight line at the loglog scale: only a few contexts are relevant to the question, while most score below 0.1 and are irrelevant. By contrast, Figure \ref{fig:analysisonV:lowskew:origin} and Figure \ref{fig:analysisonV:lowskew} show lower skewness, with a large portion of contexts clearly relevant—100\% contexts have scores above 0.1. This difference in skewness intuitively corresponds well with the query difficulty: difficult queries naturally require reasoning across more knowledge contexts, resulting in lower skewness.

\subsection{Correlation Between Skewness and Query Difficulty}
We directly quantify skewness using the area under the min-max normalized score curve: a smaller area indicates high skewness (rapid score drop-off), while a larger area reflects low skewness (many high scores and gradual decline). For example, Figure \ref{fig:analysisonV:highskew} has an area of 1.07, whereas Figure \ref{fig:analysisonV:lowskew} has 65.65.

However, query difficulty does not have explicit labels for direct assessment. Moreover, in RAG, retrieval context directly affects difficulty: clear and sufficient contexts make questions inherently easier.  Therefore, we approximate question difficulty by analyzing the scorer’s retrieval results.
In an extreme case, if the top-ranked triple context already contains the answer, the question is relatively easy, as it can be answered by matching a single fact from the KG. The scorer identifies the correct triple with high confidence during retrieval.
Conversely, if the answer-containing triple context appears lower in the ranked list, this typically indicates a more difficult, multi-hop reasoning scenario, where several semantically related but indirect triples are retrieved. Thus, we use the rank of the answer-containing triple as a proxy for query difficulty: a lower rank (i.e., further down the list) corresponds to higher difficulty.

\begin{figure}[htbp] 
    \centering
    \begin{subfigure}[b]{0.49\linewidth} 
        \centering 
        \includegraphics[width=\linewidth]{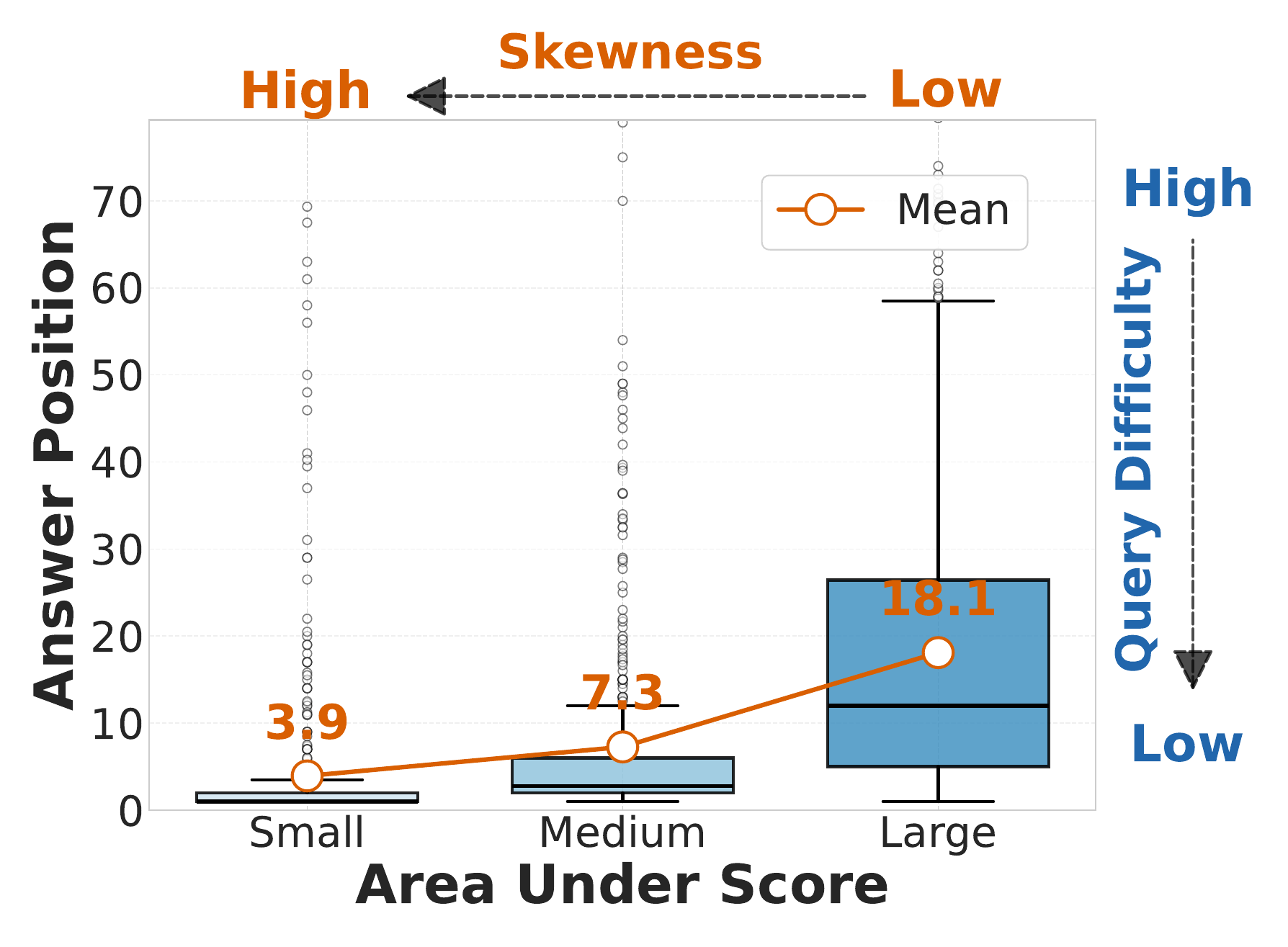}
        \caption{WebQSP Dataset}
        \label{fig:correlation:webqsp}
    \end{subfigure}
    \hfill 
    \begin{subfigure}[b]{0.49\linewidth} 
        \centering 
        \includegraphics[width=\linewidth]{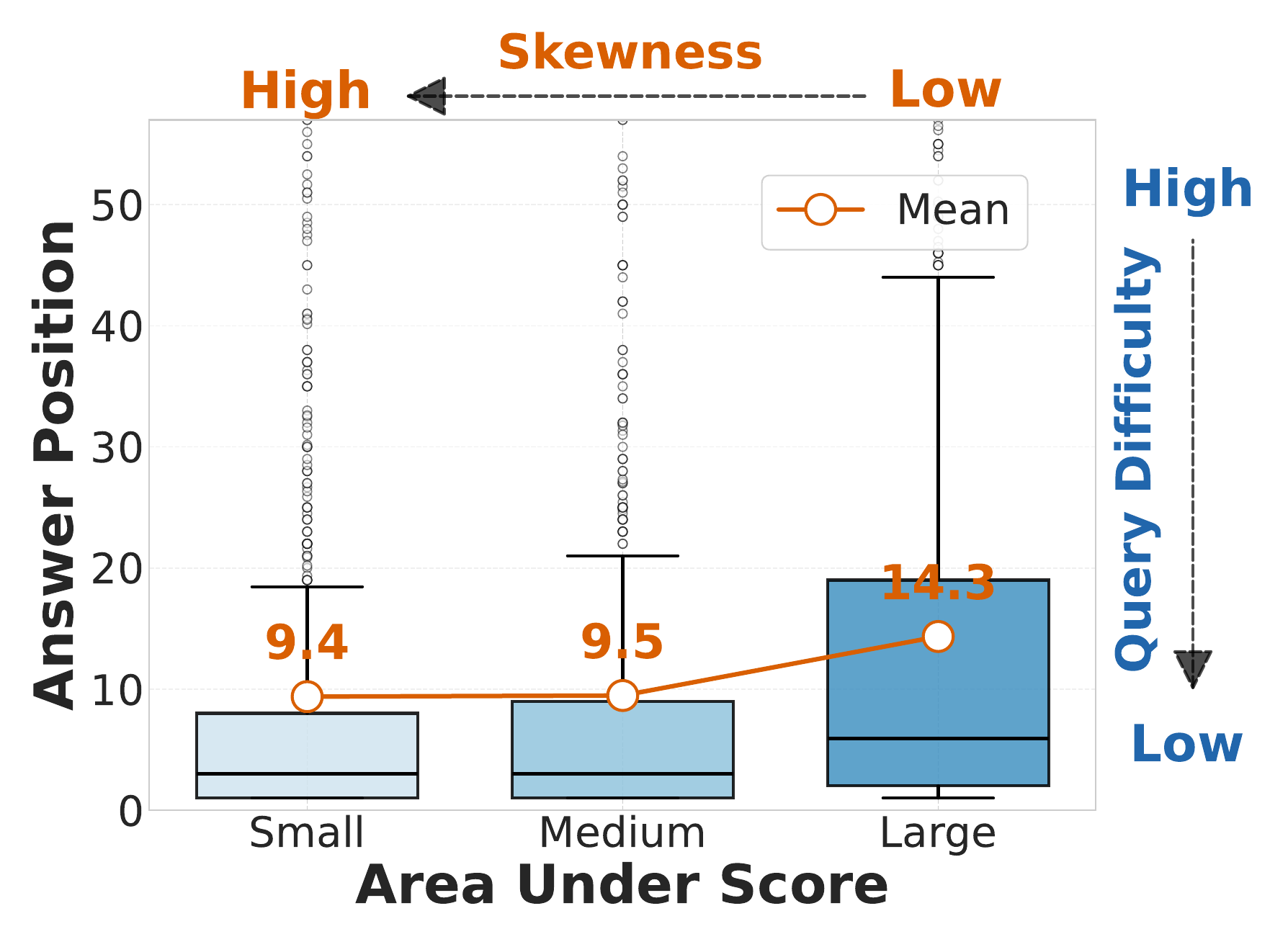}
        \caption{CWQ Dataset}   
        \label{fig:correlation:cwq}
	\end{subfigure}
    \caption{\textbf{Query Difficulty Across Score Skewness.}}
    \label{fig:answer_rank} 
\end{figure}

Using the above metrics, we examine the correlation between the skewness of retrieved scores and query difficulty. Queries are divided into three equally sized groups based on the area size, where a smaller area indicates higher skewness. As shown in Figure \ref{fig:answer_rank}, higher skewness is strongly associated with lower query difficulty, as the answer-containing context is more likely to be top-ranked. This insight allows us to classify queries as simple or difficult based on score skewness and, accordingly, route them to LLMs of different scales to balance inference cost and performance. \footnote{Further analysis is provided in Appendix \ref{appendix:cor}.}

\begin{algorithm}[ht!]
    \KwData{Dataset $\mathcal{D}$, Threshold $\theta$;}
    Initialize list of answers $\mathcal{A} \gets [\ ]$\;
    \ForEach {query $q$ in Dataset $\mathcal{D}$}{
        \If{\(\mathcal{L}(W, \theta)\) $\leq 0$}{
            $ ans \leftarrow$ \(F_S\)($I,T,q$)\;
        }
        \Else{ 
            $ ans \leftarrow$ \(F_L\)($I,T,q$)\;
        }
        Add $ans$ to $\mathcal{A}$\;
    }
    \Return{$\mathcal{A}$}\;
    \caption{Routing Framework}
    \label{alg:routing_workflow}
\end{algorithm}

\subsection{Using Skewness to Route}

In this section, we present our training-free routing framework, which can be seamlessly integrated into KG-RAG in a plug-and-play manner, as shown in Algorithm \ref{alg:routing_workflow}. By leveraging the strong correlation between score skewness of retrieved contexts and query difficulty, we require no additional training.
And any metric for measuring distribution skewness is compatible with our framework for LLM selection. 
Queries with high score skewness are routed to economical smaller LLMs, while those with low skewness are directed to more powerful larger LLMs, thereby striking an optimal balance between response quality and inference cost.

As previously mentioned, area under scores can serve as a metric of distribution skewness, yet it still has important drawbacks. 
First, it is highly sensitive to min-max normalization: differences in raw score ranges across queries lead to inconsistent scaling and unstable area comparisons. 
Second, it collapses the entire scores into a single value, sacrificing crucial information about its distribution shape. 
The consequence is that widely divergent distributions can produce similar area, undermining its ability to differentiate distinct skewness.
To address these shortcomings, we further explore new solutions to measure score skewness. 
We propose a cumulative threshold-based routing method and analyze scores from the perspective of entropy and gini coefficient to make routing decisions. 
Compared with the straightforward area under scores, these methods offer superior and robust discriminative power between simple and difficult queries. 

\textbf{Cumulative Threshold-based Routing.}
In probability and statistics, Cumulative Distribution Function \cite{kolmogoroff1933grundbegriffe, kolmogorov2018foundations} quantifies the probability a random variable $X$ is less than or equal to value $x$, offering a comprehensive view of the distribution.
Queries with high score skewness due to the rapid decay, require only a few contexts to reach a cumulative probability. 
Conversely, queries with low skewness need more contexts to achieve the same value.
Thus, cumulative threshold-based routing could guide LLM selection.
The context scores $W$ need to rank in a descending order and then normalize by the score sum into a probability distribution: $p_i = s_i / \sum_{j=1}^K s_j$. 
Building upon this, the cumulative sum $C_k$ of them is computed by: $C_k = \sum_{i=1}^{k} p_i$. 
For a given probability $P$ (e.g., 95\%), we find the smallest $k$ satisfying $C_k \ge P$. 
The function $\mathcal{L}(W, \theta)$ is defined as $k-\theta_k$.
Queries are classified as simple if $k \le \theta_k$, otherwise difficult.

\textbf{Entropy-based Routing}
Entropy \cite{6773024} can quantify the uniformity of a probability distribution. Queries with low score skewness, characterized by with more uniform and slower decaying context scores, exhibit higher entropy, while highly skewed distributions correspond to lower entropy. 
Thus, entropy can effectively assess query difficulty based on context scores, enabling dynamic routing of queries.
Calculating the entropy for scores $W$ requires converting them into a probability distribution $p = \{p_1, p_2, \dots, p_K\}$ by normalizing the scores $p_i = s_i / \sum_{j=1}^K s_j$. 
The entropy is then calculated as $H = -\sum_{i=1}^N p_i \log_2(p_i)$. 
In this context, $\mathcal{L}(W, \theta)$ refers to $H-\theta_H$.
A query is routed as simple if the entropy of context scores falls below a predetermined threshold $\theta_H$, and as difficult otherwise.

\textbf{Gini Coefficient-based Routing}
The Gini coefficient \cite{c47825a4-6703-3cd5-ae58-73cc1416e468, 1905PAmSA...9..209L}, traditionally measuring income inequality, can effectively reflect the imbalance of context scores. A higher Gini coefficient indicates greater skewness of context scores, reflecting a more pronounced imbalance among the retrieved knowledge contexts. 
Thus, Gini coefficient can serve as an effective metric for routing decision.
The calculation of the Gini coefficient for context scores $W$ involves sorting them in ascending order $s'_1 \le s'_2 \le \dots \le s'_K$ and applying the following formula: $Gini=\frac{1}{K} \left( K+1 - 2 \frac{\sum_{i=1}^K (K-i+1)s'_i}{\sum_{j=1}^K s'_j} \right)$.
Here, $\mathcal{L}(W, \theta)$ is $\theta_G-Gini$.
Queries are routed as simple if the calculated Gini coefficient exceeds a threshold $\theta_G$, indicating high skewness, and as difficult if it falls below $\theta_G$.

\section{Experiments}
\subsection{Settings}
\textbf{Datasets.}
To test the effectiveness of routing on knowledge-intensive reasoning tasks, we used two widely recognized KGQA datasets: CWQ \cite{CWQ} and WebQSP \cite{webqsp}. 
Freebase \cite{Freebase} serves as the KG for both datasets, comprising more than 120 million triples.
Query difficulty, assessed by the number of required contexts, varies across datasets. 
Specifically, WebQSP is relatively simple, containing queries with only 1 or 2 knowledge contexts. Conversely, CWQ contains more difficult reasoning problems involving up to 4 contexts.
Further details are described in Appendix \ref{appendix:datasets}.

\noindent\textbf{Baselines.}
SubgraphRAG, the latest SOTA method in KG-RAG, achieves effective and efficient performance. We thereby adopt its scorers in this study.
We compare with the following baselines: \textit{Random Routing}, which serves as the lower bound. Under the same budget, improvements over this baseline demonstrate the effectiveness of other routing methods. 
\textit{RouteLLM} \cite{ong2025routellm} and \textit{GraphRouter} \cite{feng2025graphrouter} are included as representatives of the latest routing methods. 
Following their claim of generalization across unseen data and LLMs, we use their pre-trained weights as the routing baseline against our training-free framework. 
Although these methods involve additional training cost, they still face significant challenges under the RAG setting.

\noindent\textbf{Reasoning LLMs.}
We utilize a diverse selection of open-source models for our experiments: Qwen2.5-7B-Instruct, Qwen2.5-72B-Instruct \cite{qwen2025}, Llama3.1-8B-Instruct, and Llama3.1-70B-Instruct \cite{llama3.1}. 
Based on their parameter counts, Qwen7B and Llama8B are classified as smaller LLMs, while their counterparts, Qwen72B and Llama70B, fall under the  larger LLMs.

\noindent\textbf{Evaluation Metrics.}
Consistent with prior research, we employ Hit@1 \footnote{Here, we use the officially updated evaluation code of SubgraphRAG, which leads to slightly lower results compared to those reported in the original paper (see \url{https://github.com/Graph-COM/SubgraphRAG}). However, this does not affect the validity of our routing method's evaluation.} as evaluation metrics. It measures the rate at which the correct answer appears at the top prediction position.
The objective of LLM routing is to balance performance and cost, which means a higher Hit@1 is preferable under the same call ratio of the larger LLM.
Thereby, we define \textit{Average Effectiveness} (Avg. Eff.) as the mean improvement in Hit@1 over random routing.

\subsection{Results}
\begin{table*}[t]
\small
\centering
\setlength{\tabcolsep}{3.1pt}
\begin{tabular}{llccccccc}
\toprule
Dataset & Method                   & 0\%     & 20\%                      & 40\%                      & 60\%                      & 80\%                      & 100\%    & Avg. Eff. \\
\midrule
\multirow{6}{*}{WebQSP}
  & Random Routing          & 77.52   & 78.18                     & 78.85                     & 79.51                     & 80.18                     & 80.84    & - \\
  & RouteLLM    [ICLR, 25]  & 77.52   & 78.56 (+0.38)             & 79.18 (+0.33)             & 79.55 (+0.04)             & 79.98 (-0.20)             & 80.84    & +0.14 \\
  & GraphRouter [ICLR, 25]  & 77.52   & 78.26 (+0.08)             & 79.30 (+0.45)             & 80.22 (+0.71)             & 80.71 (+0.53)             & 80.84    & +0.44 \\
  & Ours (Gini-based)       & 77.52   & \textbf{79.48 (+1.30)}    & \textit{79.98 (+1.13)}    & \textbf{81.20 (+1.69)}    & \textbf{80.96 (+0.78)}    & 80.84    & \textbf{+1.23} \\
  & Ours (Entropy-based)    & 77.52   & \textit{79.18 (+1.00)}    & \textbf{80.34 (+1.49)}    & \underline{81.08 (+1.57)} & \underline{80.77 (+0.59)} & 80.84    & \underline{+1.16} \\
  & Ours (Cumulative-based) & 77.52   & \underline{79.24 (+1.06)} & \underline{80.04 (+1.19)} & \textit{80.71 (+1.20)}    & \textit{80.71 (+0.53)}    & 80.84    & \textit{+1.00} \\
\midrule
\multirow{6}{*}{CWQ}
  & Random Routing          & 45.68   & 47.59                     & 49.51                     & 51.42                     & 53.34                     & 55.25    & - \\
  & RouteLLM    [ICLR, 25]  & 45.68   & 47.81 (+0.22)             & 50.41 (+0.90)             & 52.00 (+0.58)             & 53.61 (+0.27)             & 55.25    & +0.49 \\
  & GraphRouter [ICLR, 25]  & 45.68   & 47.83 (+0.24)             & 50.10 (+0.59)             & 52.02 (+0.60)             & 53.78 (+0.44)             & 55.25    & +0.47 \\
  & Ours (Gini-based)       & 45.68   & \underline{48.94 (+1.35)} & \textbf{50.92 (+1.41)}    & \textit{52.53 (+1.11)}    & \textbf{54.15 (+0.81)}    & 55.25    & \underline{+1.17} \\
  & Ours (Entropy-based)    & 45.68   & \textit{48.74 (+1.15)}    & \textit{50.72 (+1.21)}    & \underline{52.70 (+1.28)} & \underline{54.01 (+0.67)} & 55.25    & \textit{+1.08} \\
  & Ours (Cumulative-based) & 45.68   & \textbf{49.02 (+1.43)}    & \underline{50.89 (+1.38)} & \textbf{52.82 (+1.40)}    & \textit{53.89 (+0.55)}    & 55.25    & \textbf{+1.19} \\
\bottomrule
\end{tabular}
\caption{Routing Between Qwen2.5 Models: 7B as the Small LLM and 72B as the Large LLM.}
\label{tab:main_qwen} 
\end{table*}

Table \ref{tab:main_qwen} present the experimental results of our routing methods on the WebQSP and CWQ datasets between Qwen7B and Qwen72B.
Overall, all skewness indicators in our training-free framework consistently outperform baselines across all scenarios. 
Specifically, our gini coefficient-based, entropy-based and cumulative threshold-based methods cut larger LLM call ratio by 40\% while achieving on par with full inference of larger LLM.  
Furthermore, the gini coefficient-based routing method demonstrates a significant average effectiveness, outperforming RouteLLM by over 8x and surpassing another strong baseline, GraphRouter, by more than 2x on WebQSP dataset.
The same trend is observed on the CWQ dataset, where our method shows consistent improvements over the baselines.
These results strongly demonstrate the effectiveness of our routing methods based on score skewness to optimizing performance and cost.
Additionally, we find that the performance gap on WebQSP is smaller than that of CWQ dataset. While routing is effective in both cases, the narrower gap produces superior routing outcomes.

\begin{table*}[t]
\centering
\small
\setlength{\tabcolsep}{3.1pt}
\begin{tabular}{llccccccc}
\toprule
Dataset & Method                   & 0\%     & 20\%                      & 40\%                      & 60\%                      & 80\%                      & 100\%    & Avg. Eff. \\
\midrule
\multirow{6}{*}{WebQSP}
  & Random Routing          & 78.56   & 79.68                     & 80.80                     & 81.91                     & 83.03                     & 84.15    & - \\
  & RouteLLM    [ICLR, 25]  & 78.56   & 79.73 (+0.05)             & 80.47 (-0.33)             & 81.82 (-0.09)             & 82.86 (-0.17)             & 84.15    & -0.14 \\
  & GraphRouter [ICLR, 25]  & 78.56   & 79.67 (-0.01)             & 80.65 (-0.15)             & 81.39 (-0.52)             & \textit{83.35 (+0.32)}    & 84.15    & -0.09 \\
  & Ours (Gini-based)       & 78.56   & \underline{81.33 (+1.65)} & \underline{81.57 (+0.77)} & \textbf{82.62 (+0.71)}    & \textit{83.35 (+0.32)}    & 84.15    & \textit{+0.86} \\
  & Ours (Entropy-based)    & 78.56   & \textit{81.08 (+1.40)}    & \textbf{82.00 (+1.20)}    & \textit{82.49 (+0.58)}    & \textbf{83.66 (+0.63)}    & 84.15    & \textbf{+0.95} \\
  & Ours (Cumulative-based) & 78.56   & \textbf{81.57 (+1.89)}    & \textit{81.27 (+0.47)}    & \textbf{82.62 (+0.71)}    & \underline{83.60 (+0.57)} & 84.15    & \underline{+0.91} \\
\midrule
\multirow{6}{*}{CWQ}
  & Random Routing          & 49.90   & 51.51                     & 53.12                     & 54.72                     & 56.33                     & 57.94    & - \\
  & RouteLLM    [ICLR, 25]  & 49.90   & 51.26 (-0.25)             & 53.61 (+0.49)             & 55.03 (+0.31)             & \textit{56.75 (+0.42)}    & 57.94    & +0.24 \\
  & GraphRouter [ICLR, 25]  & 49.90   & 50.84 (-0.67)             & 52.79 (-0.33)             & 54.80 (+0.08)             & 56.16 (-0.17)             & 57.94    & -0.27 \\
  & Ours (Gini-based)       & 49.90   & \underline{52.65 (+1.14)} & \textbf{55.00 (+1.88)}    & \underline{56.16 (+1.44)} & \textbf{57.04 (+0.71)}    & 57.94    & \textbf{+1.29} \\
  & Ours (Entropy-based)    & 49.90   & \textit{52.51 (+1.00)}    & \underline{54.89 (+1.77)} & \textit{56.07 (+1.35)}    & \underline{56.78 (+0.45)} & 57.94    & \textit{+1.14} \\
  & Ours (Cumulative-based) & 49.90   & \textbf{52.68 (+1.17)}    & \textit{54.77 (+1.65)}    & \textbf{56.41 (+1.69)}    & 56.61 (+0.28)             & 57.94    & \underline{+1.20} \\
\bottomrule
\end{tabular}
\caption{Routing Between Llama3.1 Models: 8B as the Small LLM and 70B as the Large LLM.}
\label{tab:main_llama} 
\end{table*}

\begin{table*}[t]
\centering
\small
\setlength{\tabcolsep}{3.1pt} 
\begin{tabular}{llccccccc}
\toprule
Dataset & Method                   & 0\%     & 20\%                      & 40\%                      & 60\%                      & 80\%                      & 100\%    & Avg. Eff. \\
\midrule
\multirow{6}{*}{WebQSP}
  & Random Routing          & 77.52   & 78.85                     & 80.17                     & 81.50                     & 82.82                     & 84.15    & - \\
  & RouteLLM    [ICLR, 25]  & 77.52   & 78.81 (-0.04)             & 79.79 (-0.38)             & 81.33 (-0.17)             & 82.49 (-0.33)             & 84.15    & -0.23 \\
  & GraphRouter [ICLR, 25]  & 77.52   & 78.99 (+0.14)             & 80.84 (+0.67)             & 82.06 (+0.56)             & 83.42 (+0.60)             & 84.15    & +0.49 \\
  & Ours (Gini-based)       & 77.52   & \underline{81.51 (+2.66)} & \underline{82.49 (+2.32)} & \textbf{83.66 (+2.16)}    & \textbf{83.97 (+1.15)}    & 84.15    & \textbf{+2.07} \\
  & Ours (Entropy-based)    & 77.52   & 81.27 (+2.42)             & \textbf{82.74 (+2.57)}    & 83.42 (+1.92)             & \textbf{83.97 (+1.15)}    & 84.15    & \underline{+2.02} \\
  & Ours (Cumulative-based) & 77.52   & \textbf{81.63 (+2.78)}    & 82.31 (+2.14)             & \underline{83.60 (+2.10)} & \underline{83.85 (+1.03)} & 84.15    & \underline{+2.02} \\
\midrule
\multirow{6}{*}{CWQ}
  & Random Routing          & 45.68   & 48.13                     & 50.58                     & 53.04                     & 55.49                     & 57.94    & - \\
  & RouteLLM    [ICLR, 25]  & 45.68   & 48.12 (-0.01)             & 51.09 (+0.51)             & 53.58 (+0.54)             & \textit{56.05 (+0.56)}    & 57.94    & +0.40 \\
  & GraphRouter [ICLR, 25]  & 45.68   & 48.00 (-0.13)             & 50.78 (+0.20)             & 53.50 (+0.46)             & 55.31 (-0.18)             & 57.94    & +0.09 \\
  & Ours (Gini-based)       & 45.68   & \textbf{50.13 (+2.00)}    & \textbf{53.04 (+2.46)}    & \textit{54.83 (+1.79)}    & \textbf{56.53 (+1.04)}    & 57.94    & \textbf{+1.82} \\
  & Ours (Entropy-based)    & 45.68   & \textit{49.82 (+1.69)}    & \underline{52.82 (+2.24)} & \underline{54.86 (+1.82)} & \underline{56.41 (+0.92)} & 57.94    & \underline{+1.67} \\
  & Ours (Cumulative-based) & 45.68   & \underline{49.90 (+1.77)} & \textit{52.76 (+2.18)}    & \textbf{55.00 (+1.96)}    & \textit{56.05 (+0.56)}    & 57.94    & \textit{+1.62} \\
\bottomrule
\end{tabular}
\caption{Routing Across Model Families: Qwen2.5-7B-Instruct and Llama-3.1-70B-Instruct.}
\label{tab:qwen_llama}
\end{table*}

Table \ref{tab:main_llama} illustrates our experimental results routing between Llama8B and Llama70B. 
Overall, our training-free methods possess and strong flexibility and generalization, performing well across various datasets and model combinations.
Gini coefficient-based, entropy-based, and cumulative threshold-based routing methods demonstrate a significant advantage. 
Even under conditions where RouteLLM and GraphRouter fail completely, with their performance dropping below the naive random routing, our routing methods deliver a notable gain, increasing the average effectiveness by nearly 1\% on WebQSP dataset.
On the more challenging CWQ dataset, our routing methods overwhelmingly surpass all baselines. While the GraphRouter still remains ineffective, our gini coefficient-based routing method achieves average effectiveness 5x better than RouteLLM.

\begin{table*}[t]
\centering
\small
\setlength{\tabcolsep}{3.1pt} 
\begin{tabular}{llccccccc}
\toprule
Dataset & Method              & 0\%     & 20\%                      & 40\%                         & 60\%                      & 80\%                      & 100\%    & Avg. Eff. \\
\midrule
\multirow{6}{*}{WebQSP}
    & Random Routing          & 72.79     & 73.74                     & 74.68                      & 75.63                     & 76.57                     & 77.52          & - \\
		& RouteLLM    [ICLR, 25]  & 72.79     & 74.26 (+0.52)             & 75.25 (+0.57)              & 76.17 (+0.54)             & 77.33 (+0.76)             & 77.52          & +0.60 \\
		& GraphRouter [ICLR, 25]  & 72.79     & 73.59 (-0.15)             & 74.69 (+0.01)              & 75.92 (+0.29)             & 77.03 (+0.46)             & 77.52          & +0.15 \\
		& Ours (Gini-based)       & 72.79     & \textit{75.12 (+1.38)}    & \textbf{77.27 (+2.59)}     & \textit{76.78 (+1.15)}    & \textbf{77.52 (+0.95)}    & 77.52          & \textit{+1.52} \\
		& Ours (Entropy-based)    & 72.79     & \textbf{75.31 (+1.57)}    & \underline{77.21 (+2.53)}  & \underline{77.21 (+1.58)} & \underline{77.46 (+0.89)} & 77.52          & \underline{+1.64} \\
		& Ours (Cumulative-based) & 72.79     & \underline{75.25 (+1.51)} & \underline{77.21 (+2.53)}  & \textbf{77.58 (+1.95)}    & \textit{77.40 (+0.83)}    & 77.52          & \textbf{+1.71} \\
\midrule
\multirow{6}{*}{CWQ}
  & Random Routing          & 41.18   & 42.08                         & 42.98                      & 43.88                     & 44.78                     & 45.68    & - \\
  & RouteLLM    [ICLR, 25]  & 41.18   & 41.97 (-0.11)                 & 42.65 (-0.33)              & 43.39 (-0.49)             & \textit{44.41 (-0.37)}    & 45.68    & -0.33 \\
  & GraphRouter [ICLR, 25]  & 41.18   & 41.49 (-0.59)                 & 42.03 (-0.95)              & 43.16 (-0.72)             & 44.72 (-0.06)             & 45.68    & -0.58 \\
  & Ours (Gini-based)       & 41.18   & \underline{42.45 (+0.37)}     & \underline{43.59 (+0.61)}  & \textbf{44.94 (+1.06)}    & \underline{45.37 (+0.59)} & 45.68    & \underline{+0.66} \\
  & Ours (Entropy-based)    & 41.18   & \textbf{42.65 (+0.57)}        & \textbf{43.64 (+0.66)}     & \textbf{44.94 (+1.06)}    & \textit{45.20 (+0.42)}    & 45.68    & \textbf{+0.68} \\
  & Ours (Cumulative-based) & 41.18   & \textit{42.31 (+0.23)}        & \textit{43.50 (+0.52)}     & \textit{44.86 (+0.98)}    & \textbf{45.51 (+0.73)}    & 45.68    & \textit{+0.62} \\
\bottomrule
\end{tabular}
\caption{Routing of Retrieved Contexts Nums (Top-K, k=10 vs.k=100) of Qwen2.5-7B-Instruct.}
\label{tab:dynamic}
\end{table*}

\subsection{Case Study}
\subsubsection{Routing Across Diverse Model Sizes}
\begin{figure}[htbp] 
	\centering 
	\begin{subfigure}[b]{0.493\linewidth} 
    \includegraphics[width=0.998\linewidth]{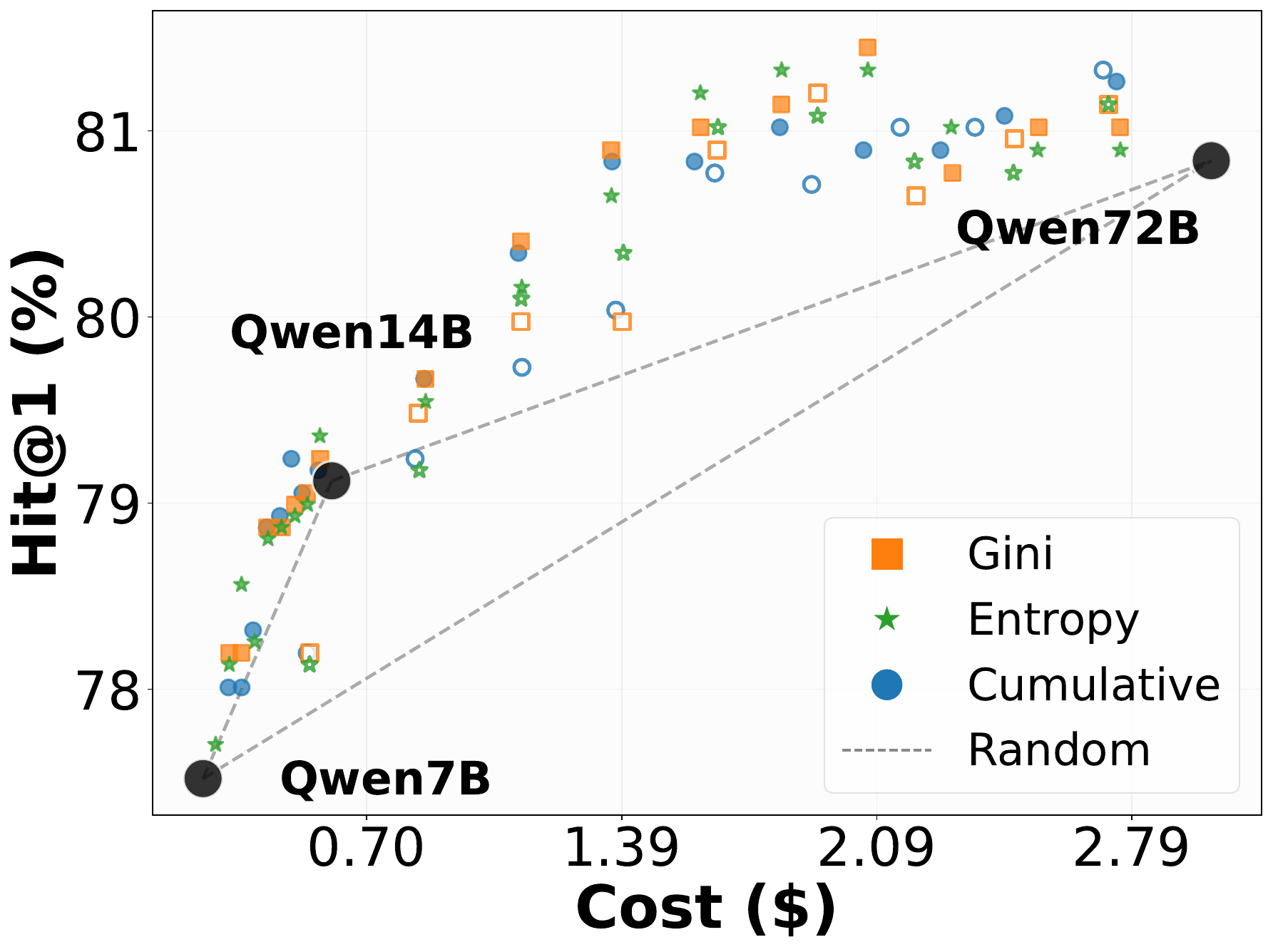} 
    \caption{\textbf{ WebQSP Dataset.}}
    \label{fig:multi_webqsp}
	  \end{subfigure}
	\hfill 
	\begin{subfigure}[b]{0.493\linewidth}
    \centering
      \includegraphics[width=0.998\linewidth]{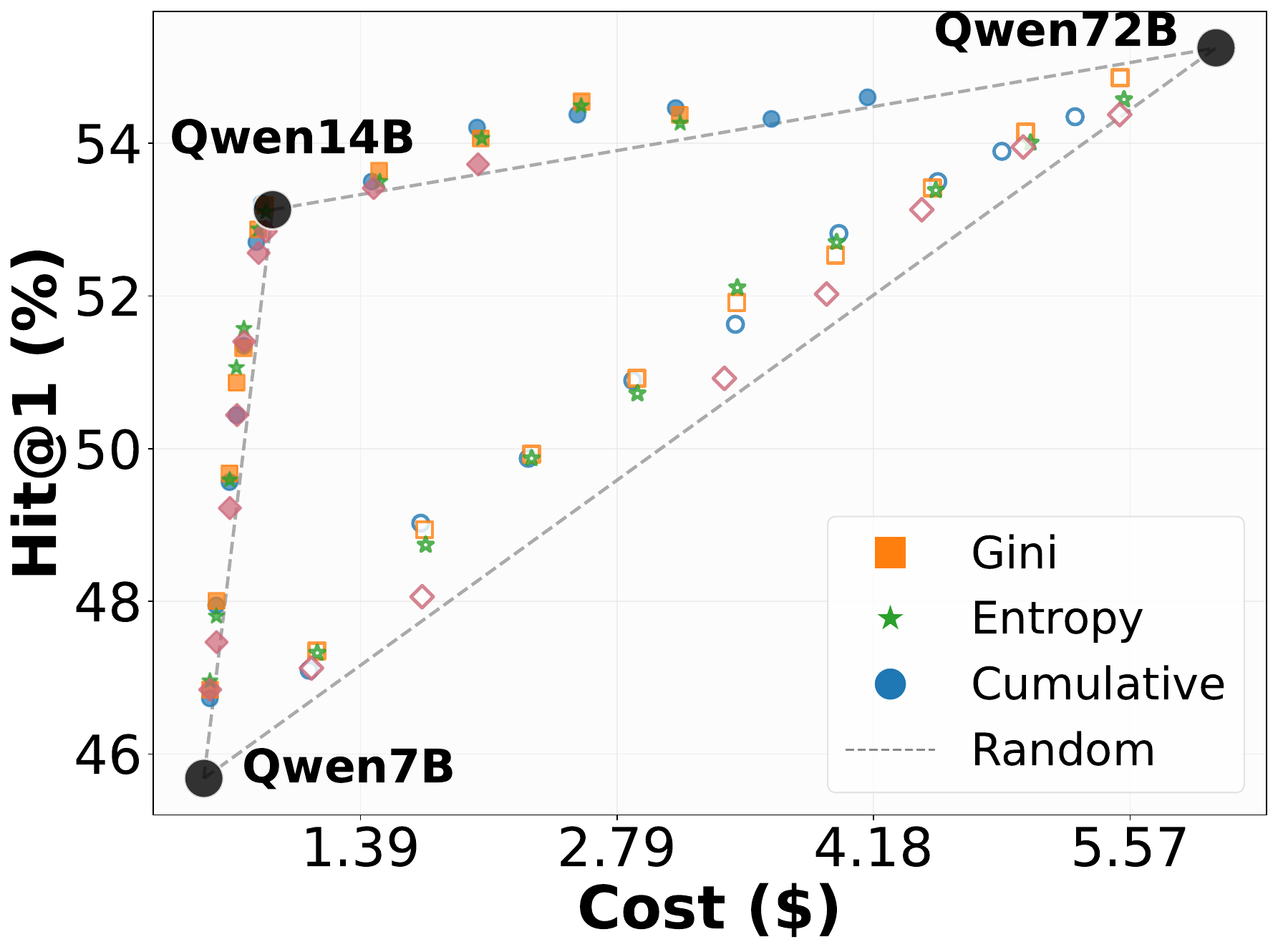} 
      \caption{\textbf{ CWQ Dataset.}}
      \label{fig:multi_cwq}
	    \end{subfigure}
	  \caption{Routing Between Multiple Models}
    \label{fig:multi} 
\end{figure}
In this section, we further extend our methods to include an additional model size to demonstrate its generalization. Specifically, we use Qwen7B as the small LLM, Qwen72B as the large LLM, and Qwen14B as the medium LLM, which sits between them in terms of parameters, performance, and cost. The experimental results are presented in Figure \ref{fig:multi}. Notably, including the medium LLM significantly improves the trade-off between performance and cost. Routing with the medium model yields much better results than routing directly between the small and large models. Overall, our method also demonstrates clear effectiveness, consistently outperforming the random routing baseline. This highlights the generalization and robustness of our methods, effectively enabling training-free routing across multiple LLM sizes and allowing for finer-grained cost-performance trade-offs tailored to the user's budget.

\subsubsection{Routing Across Model Families}
As shown in Table \ref{tab:qwen_llama}, we further explore routing across different model families, such as between Qwen7B and Llama70B, which are trained on different corpora and have different model architectures, to evaluate the generalization ability of our routing methods. Even when routing across distinct model families, our routing framework remains consistently effective, demonstrating strong robustness and generalization. 
Specifically, on the WebQSP dataset, the average effectiveness of the gini coefficient-based routing method is over 4x higher than GraphRouter. 
On the CWQ dataset, our routing methods significantly outperform all baselines. 
At a 40\% large LLM call ratio, our gini coefficient-based routing method brings a 2.57\% and 2.46\% improvement over random routing on the WebQSP and CWQ datasets, respectively.
These results highlight the superiority of our routing methods in reducing inference costs while maintaining inference performance.
Our gini coefficient-based routing method achieves the best average effectiveness. Therefore, we recommend it for routing across model families.

\subsubsection{Routing of Retrieved Contexts Nums}
In addition to routing between models of different sizes, dynamically adjusting the number of retrieved contexts is another promising direction. For simple queries, reducing the number of retrieved contexts effectively decreases the input token count and lowers inference costs. In this case study, we employ our skewness-based routing algorithm to achieve this: for complex queries, we retain 100 retrieved contexts as before, whereas for simple queries, only 10 contexts are used. 
As illustrated in Table~\ref{tab:dynamic}, our routing methods demonstrate remarkable efficiency and performance. Notably, even when using only 10 retrieved contexts for 60\% of the queries, our cumulative threshold-based routing method outperformes using 100 contexts for all queries on WebQSP, while drastically reducing inference costs. Furthermore, the average effectiveness of this method is nearly 3x that of RouteLLM. 
This highlights the broad applicability of our proposed skewness-based plug-and-play algorithm and its promising potential for future applications.

\subsubsection{Other Scorer in KG-RAG}
The focus of this work is not on improving the scorers used in the KG-RAG pipeline. Instead, our method leverages the score distributions produced by existing scorers in a plug-and-play manner for routing, making it largely agnostic to the choice of scorer.
To further demonstrate the generality of our method, we conduct additional evaluations using a domain fine-tuned MLP scorer in conjunction with the \texttt{gte-large-en-v1.5} embedding model as new scorer. 
Across all settings, our routing method still achieves substantial performance gains over existing routing methods. These results provide further validation of the effectiveness and broad applicability of our methods, as shown in the Table \ref{tab:other_scorer}.
Although the scorer trained on the semantic similarity between context and query don't perform as well as SubgraphRAG, our routing methods still lead the pack, a strong testament to their versatility. Specifically, our entropy-based routing method is particularly notable, with its average effectiveness being nearly 3x that of RouteLLM.

\begin{table*}[t]
\centering
\small
\setlength{\tabcolsep}{3.1pt} 
\begin{tabular}{llccccccc}
\toprule
Dataset & Method            & 0\%       & 20\%                      & 40\%                      & 60\%                        & 80\%                        & 100\%    & Avg. Eff. \\
\midrule
\multirow{6}{*}{WebQSP}
  & Random Routing          & 75.61     & 76.42                     & 77.23                     & 78.05                       & 78.86                       & 79.67    & - \\
  & RouteLLM    [ICLR, 25]  & 75.61     & 76.66 (+0.24)             & 77.40 (+0.17)             & 78.01 (-0.04)               & 78.50 (-0.36)               & 79.67    & +0.003 \\
  & GraphRouter [ICLR, 25]  & 75.61     & 76.84 (+0.42)             & 77.95 (+0.72)             & 78.93 (+0.88)               & 80.04 (+1.18)               & 79.67    & +0.80 \\
  & Ours (Gini-based)       & 75.61     & \underline{77.46 (+1.04)} & \underline{78.62 (+1.39)} & \underline{79.55 (+1.50)}   & \underline{79.67 (+0.81)}   & 79.67    & \textit{+1.19} \\
  & Ours (Entropy-based)    & 75.61     & \textit{77.40 (+0.98)}    & \textbf{78.69 (+1.46)}    & \underline{79.55 (+1.50)}   & \textbf{79.73 (+0.87)}      & 79.67    & \underline{+1.20} \\
  & Ours (Cumulative-based) & 75.61     & \textbf{77.58 (+1.16)}    & \textit{78.56 (+1.33)}    & \textbf{79.79 (+1.74)}      & \textit{79.61 (+0.75)}      & 79.67    & \textbf{+1.25} \\
\midrule
\multirow{6}{*}{CWQ}
  & Random Routing          & 43.98     & 45.72                     & 47.46                     & 49.20                     & 50.94                     & 52.68          & - \\
  & RouteLLM    [ICLR, 25]  & 43.98     & 46.13 (+0.41)             & 48.12 (+0.66)             & 49.70 (+0.50)             & 51.29 (+0.35)             & 52.68          & +0.48 \\
  & GraphRouter [ICLR, 25]  & 43.98     & 45.71 (-0.01)             & 47.78 (+0.32)             & 49.67 (+0.47)             & 51.29 (+0.35)             & 52.68          & +0.28 \\
  & Ours (Gini-based)       & 43.98     & \underline{46.84 (+1.12)} & \underline{49.05 (+1.59)} & \textit{50.89 (+1.69)}    & \textbf{51.97 (+1.03)}    & 52.68          & \underline{+1.36} \\
  & Ours (Entropy-based)    & 43.98     & \textbf{46.87 (+1.15)}    & \textbf{49.16 (+1.70)}    & \textbf{51.06 (+1.86)}    & \textit{51.85 (+0.91)}    & 52.68          & \textbf{+1.41} \\
  & Ours (Cumulative-based) & 43.98     & \textit{46.45 (+0.73)}    & \textit{48.99 (+1.53)}    & \underline{51.01 (+1.81)} & \underline{51.91 (+0.97)} & 52.58          & \textit{+1.26} \\
\bottomrule
\end{tabular}
\caption{Routing Between Qwen Models with Other Scorer: 7B as the Small LLM and 72B as the Large LLM.}
\label{tab:other_scorer}
\end{table*}

\subsubsection{Computational Efficiency of Routing}
Table \ref{tab:time_results} presents a runtime comparison of different routing methods.
Our method is training-free and extremely fast, as it estimates the skewness level simply by analyzing the score distribution of the top-$K$ retrieved contexts.
In contrast, training-based methods like GraphRouter and RouteLLM require GPU resources for deployment and suffer significant slowdowns when only CPUs are available. Our method, however, runs efficiently on CPUs alone.
Specifically, GraphRouter incurs substantial latency due to query embedding, running 5,460× slower than our cumulative-based method on CPUs, and still 2,714× slower even with GPU acceleration.

\begin{table*}[htbp]
    \small
    \centering
    \setlength{\tabcolsep}{3.8pt}
    \begin{tabular}{l|c|c|c|c|c|c}
    \toprule
    Model    & RouteLLM & GraphRouter  & Our (Gini-based)   & Our (Entropy-based)  & Our (Cumulative-based)  & Our (Area-based)  \\
    \midrule
    CPU      & 25.21  & 46.41        & 0.01         & 0.03            & 0.01             & 0.01 \\
    GPU      & 10.12  & 23.07        & -            & -               & -                & - \\
    \bottomrule
    \end{tabular}
    \caption{Time Statistics of Different Routing Methods (ms/query).}
    \label{tab:time_results}
\end{table*}

\section{Related Works}
In LLM routing, the necessity of maximizing quality under cost constraints is challenging.
Numerous studies have sought to mitigate this issue by training a model on a specifically constructed dataset designed to determine the most suitable LLM to handle the query.
For instance, AutoMix \cite{aggarwal2024automix} strategically routes queries to large LLMs, employing a POMDP router based on answer confidence of small LLMs. However, it requires multiple model calls, increasing latency and overhead.
Hybrid-LLM \cite{ding2024hybrid} trains a BERT-based router that performs binary classification based on the predicted query difficulty.
RouteLLM \cite{ong2025routellm} leverages human preference data and employs data augmentation techniques to enhance the router performance.
Afterwards, GraphRouter \cite{feng2025graphrouter} builds router for LLM selections from the graph perspective and models the LLM selection problem as an edge prediction task.
Compared to them, our methods offer a novel solution that does not necessitate training on specific datasets. This not only saves substantial data synthesis and training costs but also ensures high flexibility and generalization, particularly in scenarios like KG-RAG where dedicated training data is infeasible.

Knowledge Graph Question Answering (KGQA) serves as a crucial application for evaluating KG-RAG, which requires precise answers from structured KGs.
These KG-RAG methods are primarily divided into two categories: decoupled retrieval methods and integrated retrieval methods based on whether LLMs dynamically participate in the retrieval process on KGs.
The integrated retrieval methods \cite{jiang2024kg, sun2024thinkongraph, ma2024debategraphflexiblereliable, ma2025thinkongraph} exemplified by ToG, enable the LLMs to interactively explore related entities and relations on KGs and perform reasoning based on the retrieved knowledge.
However, this tight coupling requires frequent LLM calls, leading to substantial inference overhead.
The decoupled retrieval methods \cite{mavromatis2024gnnraggraphneuralretrieval, luo2024reasoning, he2024gretriever, hu2024graggraphretrievalaugmentedgeneration, wang2025pathpoolingtrainfreestructure, li2025simple} focus on distilling prior knowledge from KGs into LLMs. 
For instance, RoG fine-tunes LLMs to generate relation paths as faithful plans that are then used to retrieve valid paths from KGs.
However, this method incurs substantial computational expense due to fine-tuning and processing latency, which can amount to thousands of seconds per query.
In contrast, SubgraphRAG, demonstrates superior effective and efficient results by integrating a lightweight multilayer perceptron (MLP) with structural and semantic information. 
Accordingly, all our subsequent experiments are conducted based on it.

\section{Conclusion}
In this paper, we present the first LLM routing framework tailored for KG-RAG in a plug-and-play manner. This is motivated by the observation that the skewness of retrieval score distributions naturally reflects query difficulty in reasoning over external knowledge, aligning well with the RAG paradigm. By exploiting this skewness, our method dynamically routes queries among LLMs of different scales, achieving an effective trade-off between performance and cost. This simple yet robust solution also overcomes prior methods' dependence on costly training data and generalizes well across datasets and model sizes. Extensive experiments validate the effectiveness of our method. 
We offer a new perspective for efficient RAG deployment by reducing costs through LLM routing. Future work will focus on extending this training-free routing paradigm to broader RAG settings, including text-based and multimodal scenarios.

\section*{Limitations}
SkewRoute is a highly flexible and generalizable method based on score skewness. However, current experiments focus mainly on KG-RAG, while its principles should also apply to broader RAG settings, such as chunk-based text RAG or multimodal RAG. Consequently, exploring SkewRoute’s effectiveness across these wider RAG scenarios is a promising direction. We acknowledge this as a limitation of the current work and leave it for future study. 

\bibliography{custom}

\clearpage
\appendix
\section{Appendix}
\label{sec:appendix}
\subsection{Datasets}
\label{appendix:datasets}
The WebQuestionsSP (WebQSP) dataset, extends the original WebQuestions dataset by providing 4,737 natural language questions annotated with corresponding SPARQL queries executable against Freebase.
The ComplexWebQuestions (CWQ) dataset is designed to assess models' abilities to answer complex queries that necessitate reasoning across multiple knowledge contexts of KGs.  
WebQSP and CWQ are widely utilized for training and evaluating KG-RAG methods. 

Table \ref{tab:datasets} summarizes the statistics of the datasets used in our experiments. 
Specifically, the WebQSP dataset consists of 2,826 training samples and 1,628 test samples, with a maximum hop of 2.
The CWQ dataset contains 27,639 training samples and 3,531 test samples, with a maximum hop count of 4. 
\begin{table}[htbp]
    \centering
    \begin{tabular}{@{}c|ccc@{}}
        \toprule
        Datasets & Train  & Test  & Max hop \\
        \midrule
        CWQ      & 27,639 & 3,531 & 4       \\
        WebQSP   & 2,826  & 1,628 & 2       \\
        \bottomrule
    \end{tabular}
    \caption{Statistics of Datasets.}
    \label{tab:datasets}
\end{table}

Table \ref{tab:hop_dist} details the distribution of query hops across the datasets. 
The WebQSP dataset is relatively simpler, comprising only 1-hop and 2-hop queries, with 1-hop queries forming the majority. In contrast, CWQ is a more complex dataset. 
Although it comprises 1-hop and 2-hop queries, a considerable percentage (20.8\%) consists of queries requiring three or four hops.
\begin{table}[htbp]
    \centering
    \begin{tabular}{@{}cccc@{}}
        \toprule
        Dataset & 1 hop                         & 2 hop   & $\geq$ 3 hop \\ \midrule
        WebQSP  & 65.5                      \% & 34.5\% & 0.00\%       \\
        CWQ     & 40.9                      \% & 38.3\% & 20.8\%      \\ \bottomrule
    \end{tabular}
    \caption{Statistics of Query Hops of Datasets}
    \label{tab:hop_dist}
\end{table}

\subsection{Implementation Details.}
For reasoning, we use one-shot prompting to LLMs' generation. Consistent with SubgraphRAG, the prompt template we employe is detailed in Appendix \ref{appendix:prompt}. The temperature parameter is set to 0 for reproduction, and the maximum token length for generation is fixed at 4000.

\subsection{Performance and Cost}
\label{appendix:performance}
The performance of the KG-RAG state-of-the-art method SubgraphRAG on the CWQ and WebQSP datasets is presented in Table \ref{tab:rank_results} with 100 retrieved triples.
\begin{table}[htbp]
    \small
    \centering
    \setlength{\tabcolsep}{4pt}
    \begin{tabular}{l|c|c|c|c}
        \toprule
        Models & Llama-8b & Llama-70b & Qwen-7b & Qwen-72b \\
        \midrule
        \multicolumn{5}{c}{\textbf{CWQ}} \\
        \midrule
        F1-Score & 46.83 & 53.53 & 42.77 & 52.11 \\
        Hit@1    & 49.90 & 57.94 & 45.68 & 55.25 \\
        \midrule
        \multicolumn{5}{c}{\textbf{WebQSP}} \\
        \midrule
        F1-Score & 69.29 & 73.93 & 67.55 & 70.76 \\
        Hit@1    & 78.56 & 84.15 & 77.52 & 80.84 \\
        \bottomrule
    \end{tabular}
    \caption{Results of SubgraphRAG with 100 Triples on CWQ and WebQSP datasets.}
    \label{tab:rank_results}
\end{table}
All our experiments are conducted on the LLM cloud service platform.
Table \ref{tab:rank_results} presents the cost per one million tokens of different LLMs. More pricing details are provided on the official Silicon Flow website\footnote{\url{https://siliconflow.cn/}}.
\begin{table}[htbp]
    \scriptsize
    \centering
    \setlength{\tabcolsep}{1.2pt}
    \begin{tabular}{l|c|c|c|c|c|c}
        \toprule
        Models             & Qwen7b & Qwen14b & Qwen32b & Qwen72b & Llama8b &Llama70b \\
        \midrule
        Cost/M tokens      & 0.0485   & 0.0970    &0.1746   & 0.5724    & 0.0485    & 0.5724 \\
        \bottomrule
    \end{tabular}
    \caption{Inference Cost (\$) of LLMs with Different Scale on SiliconFlow.}
    \label{tab:cost_results}
\end{table}
Driven by architectural innovations and scaling laws, LLMs has emerged.
As shown in the Table \ref{tab:rank_results}, the cost-efficient smaller LLMs have limited capabilities while powerful larger LLMs incur high inference cost. 
This inherent performance-cost trade-off poses a significant deployment challenge. 
LLM routing offers an effective solution by directing queries to balance response quality with cost efficiency. 
For users, routing enables them handling simpler queries on their devices locally while call to expensive cloud APIs only for more difficult queries.
For LLM platform providers, routing facilitates automatically directing simple queries to lower-cost smaller LLMs on the backend without sacrificing user experience. 
Consequently, LLM routing provides a flexible and cost-effective method, helping to unlock the full potential of LLMs across various scales while accommodating diverse budget and performance requirements.

\subsection{Score Skewness on WebQSP Dataset}
\begin{figure}[htbp] 
    \centering 
    \begin{subfigure}[b]{0.49\linewidth} 
        \centering 
        \includegraphics[width=\linewidth]{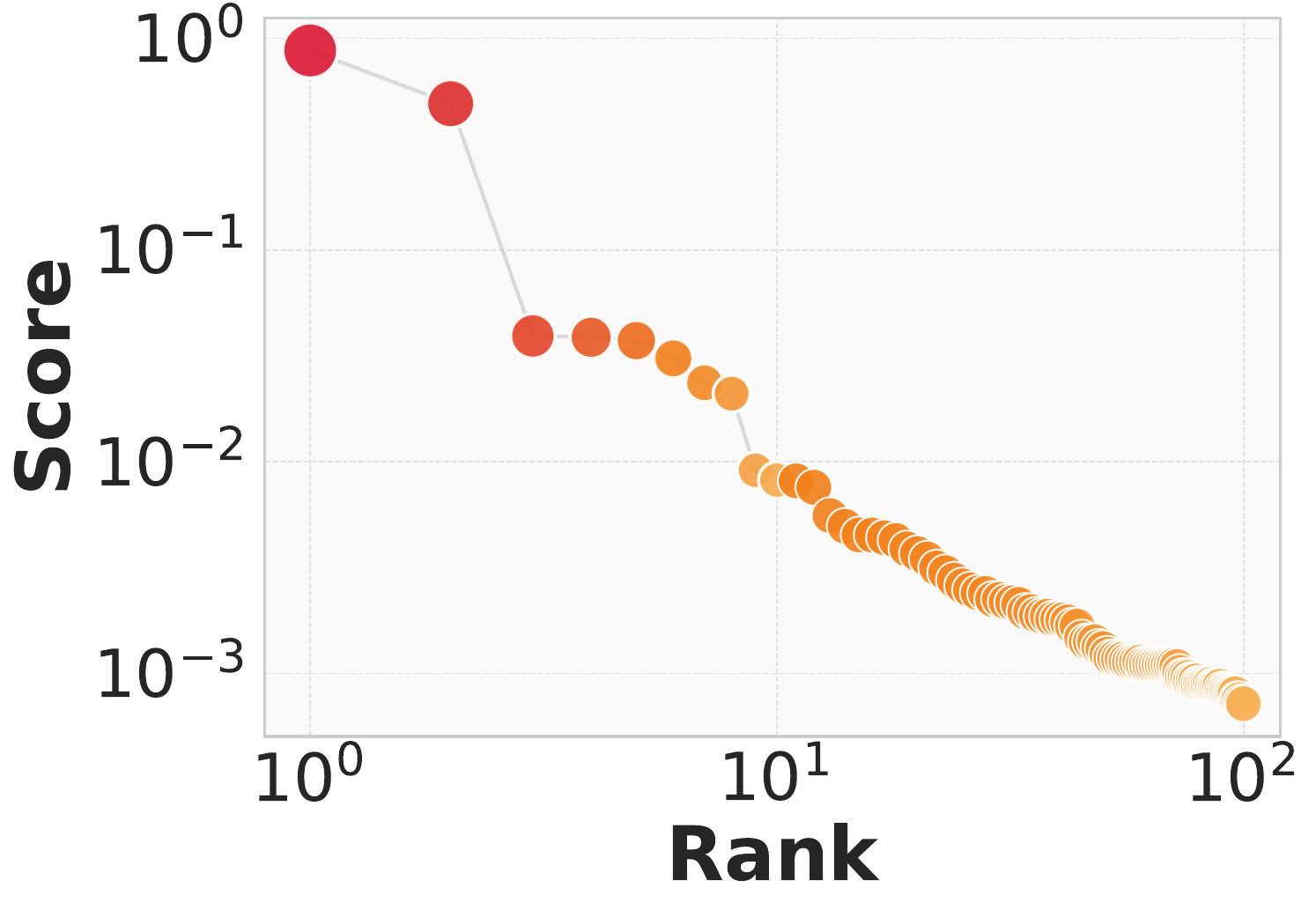}
        \caption{High Skewness}
        \label{fig:analysison:highskew}
    \end{subfigure}
    \hfill 
    \begin{subfigure}[b]{0.49\linewidth}
        \centering 
        \includegraphics[width=\linewidth]{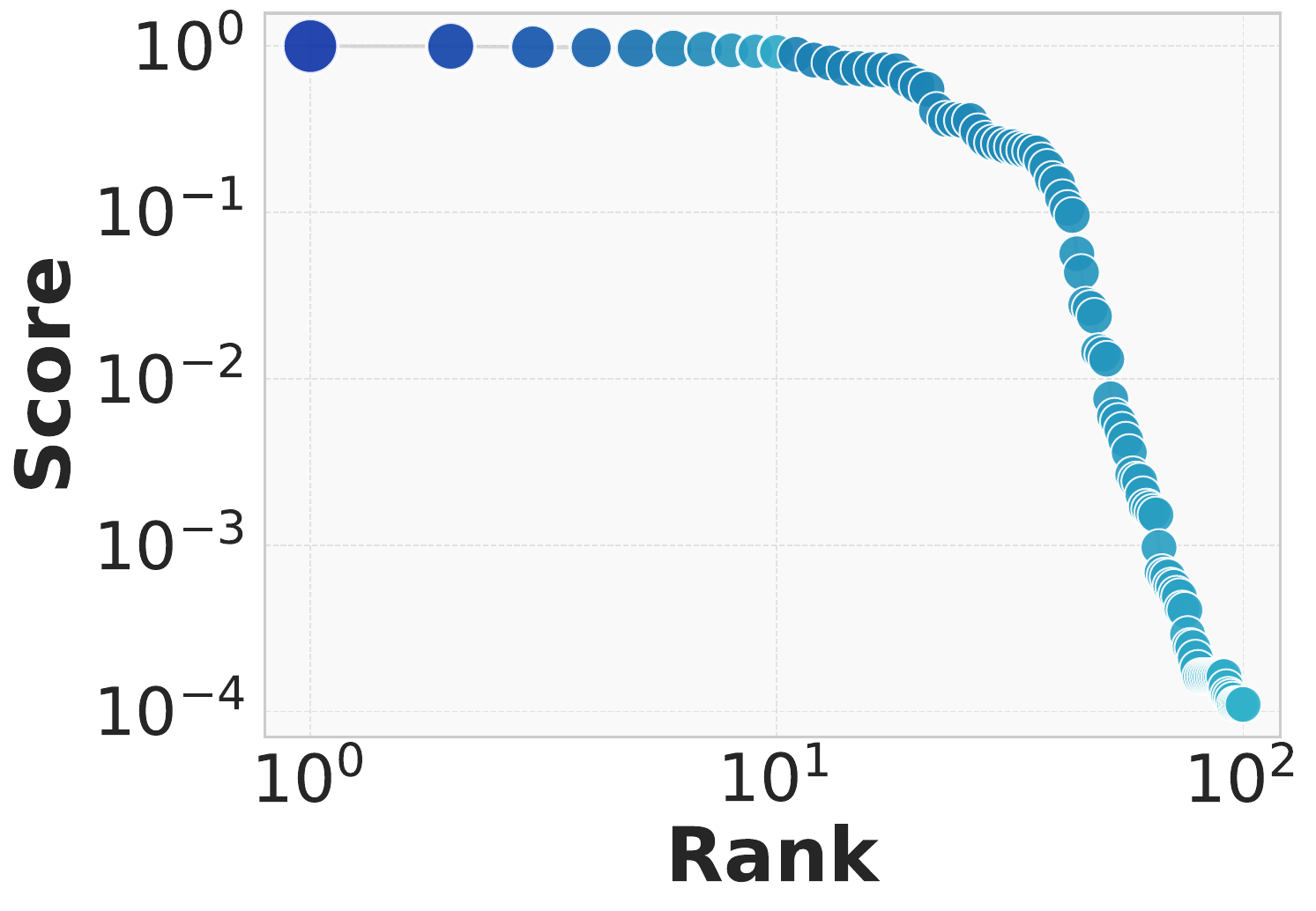}
        \caption{Low Skewness}
        \label{fig:analysison:lowskew}
    \end{subfigure}
    \caption{\textbf{Score of Retrieved Contexts in WebQSP.}}
    \label{fig:analysison} 
\end{figure}

Figure \ref{fig:analysison} illustrates the scores of retrieved contexts for queries within the WebQSP dataset, a trend likewise observed in the CWQ dataset as described earlier.
Specifically, the left panel displays high skewness, with retrieved context scores declining sharply as rank increases.
This highly skewed scores indicate that a few of high-ranked contexts contribute the vast majority of the total score, while the extensive long tail comprises numerous contexts with significantly lower individual scores.
In stark contrast, the right panel shows low skewness, deviating from this linear drop pattern.
A prominent inflection point in its upper-right region indicates that contexts, particularly those of higher scores, diminish at a substantially more gradual pace than that of high skewness.

\subsection{Cumulative Probability of Context Scores}
\begin{figure}[t] 
    \centering 
    \begin{subfigure}[b]{0.49\linewidth} 
        \centering 
        \includegraphics[width=\linewidth]{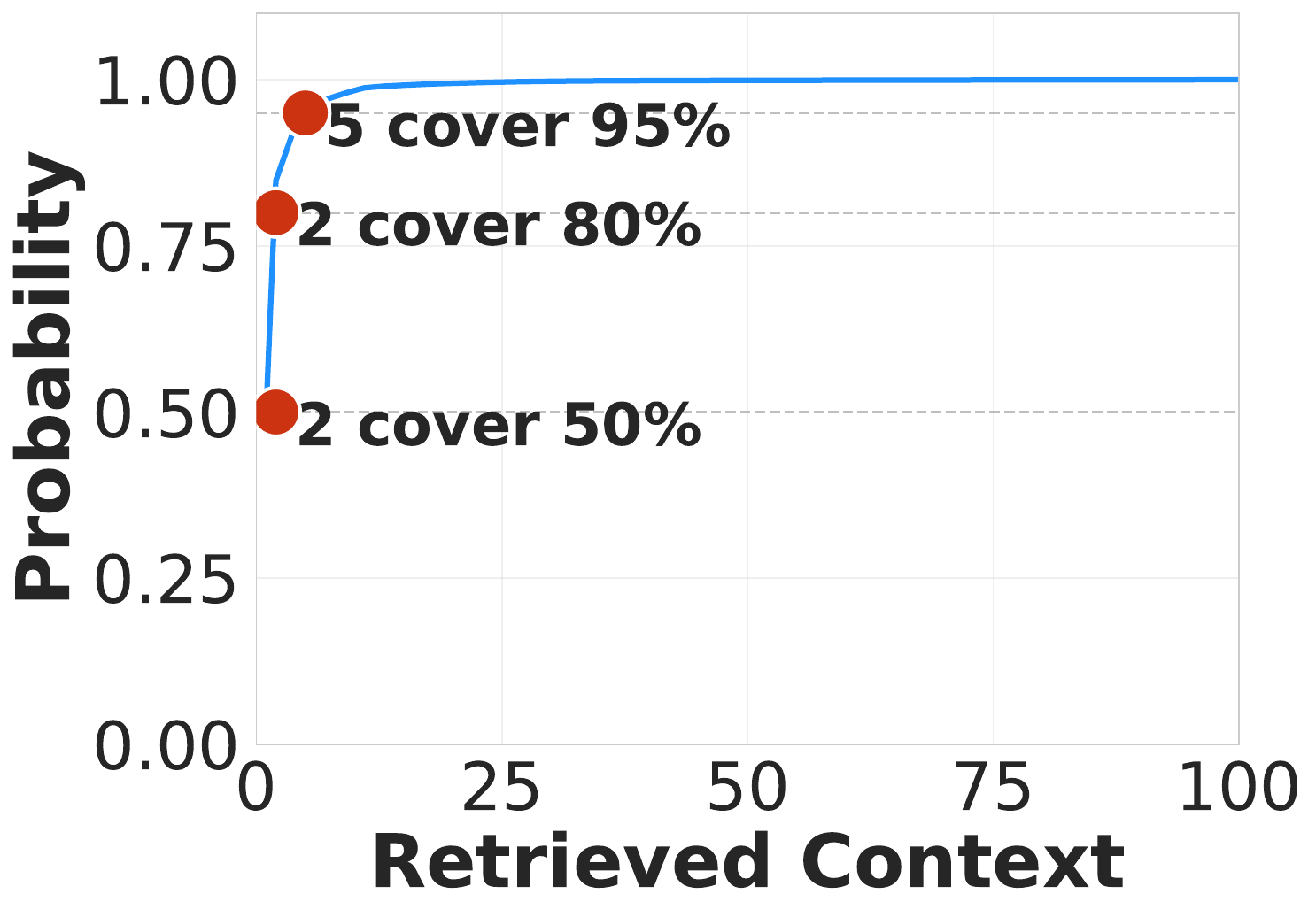}
        \caption{High Skewness} 
        \label{fig:cdf_webqsp_high} 
    \end{subfigure}
    \hfill 
    \begin{subfigure}[b]{0.49\linewidth}
        \centering
        \includegraphics[width=\linewidth]{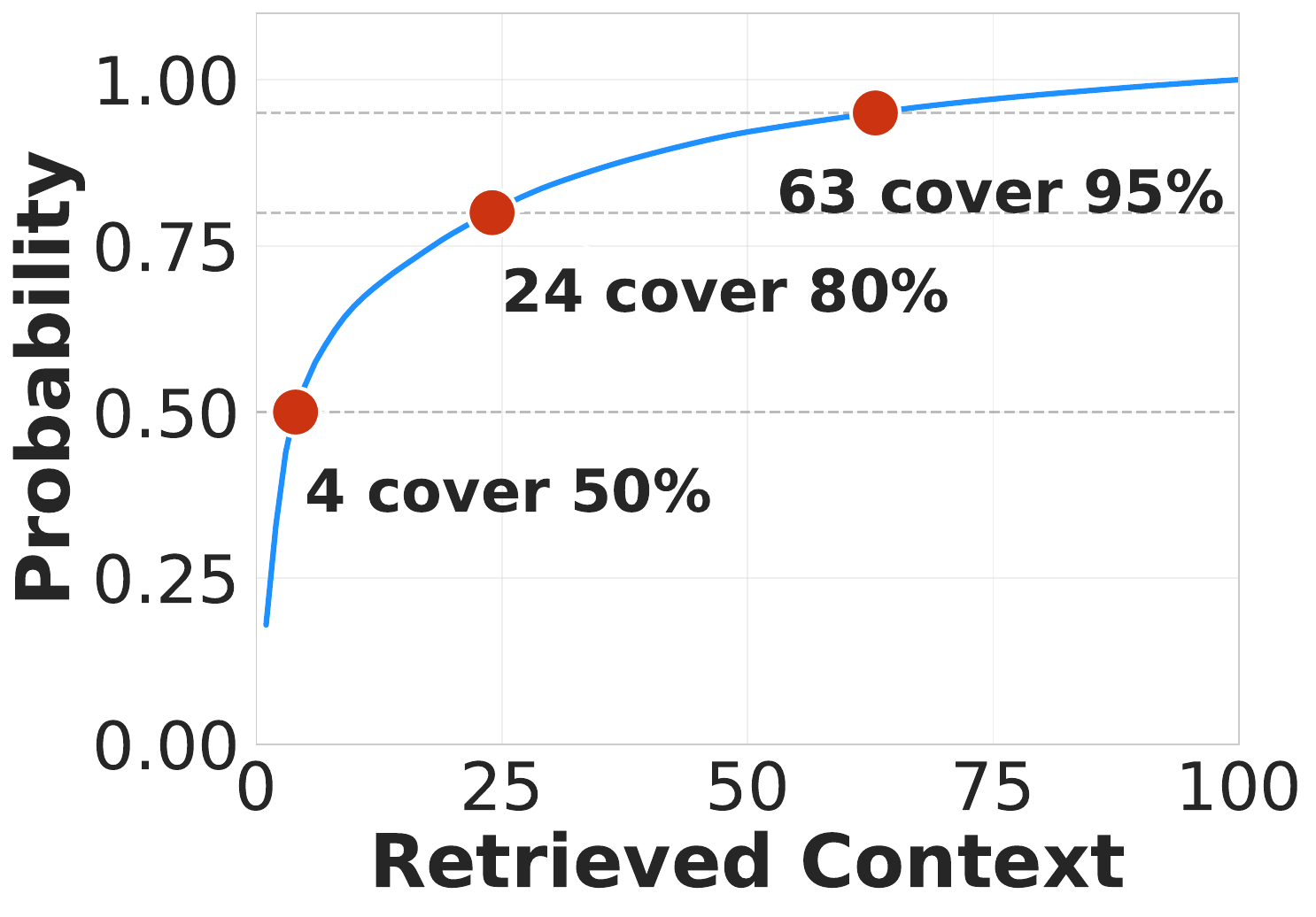}
        \caption{Low Skewness}
        \label{fig:cdf_webqsp_low}
    \end{subfigure}
    \begin{subfigure}[b]{0.49\linewidth}
        \centering
        \includegraphics[width=\linewidth]{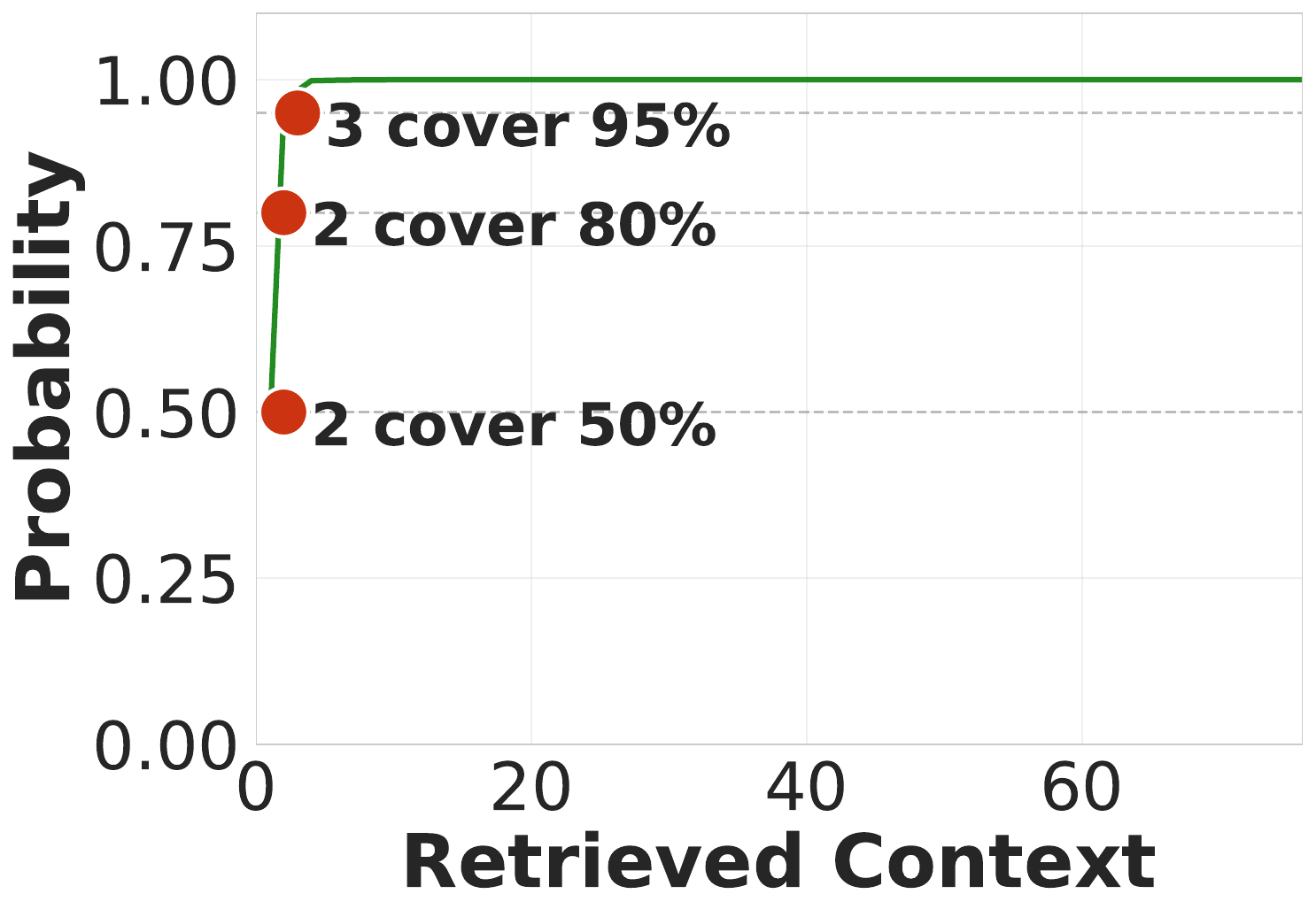}
        \caption{High Skewness}
        \label{fig:cdf_cwq_high}
    \end{subfigure}
    \hfill 
    \begin{subfigure}[b]{0.49\linewidth}
        \centering
        \includegraphics[width=\linewidth]{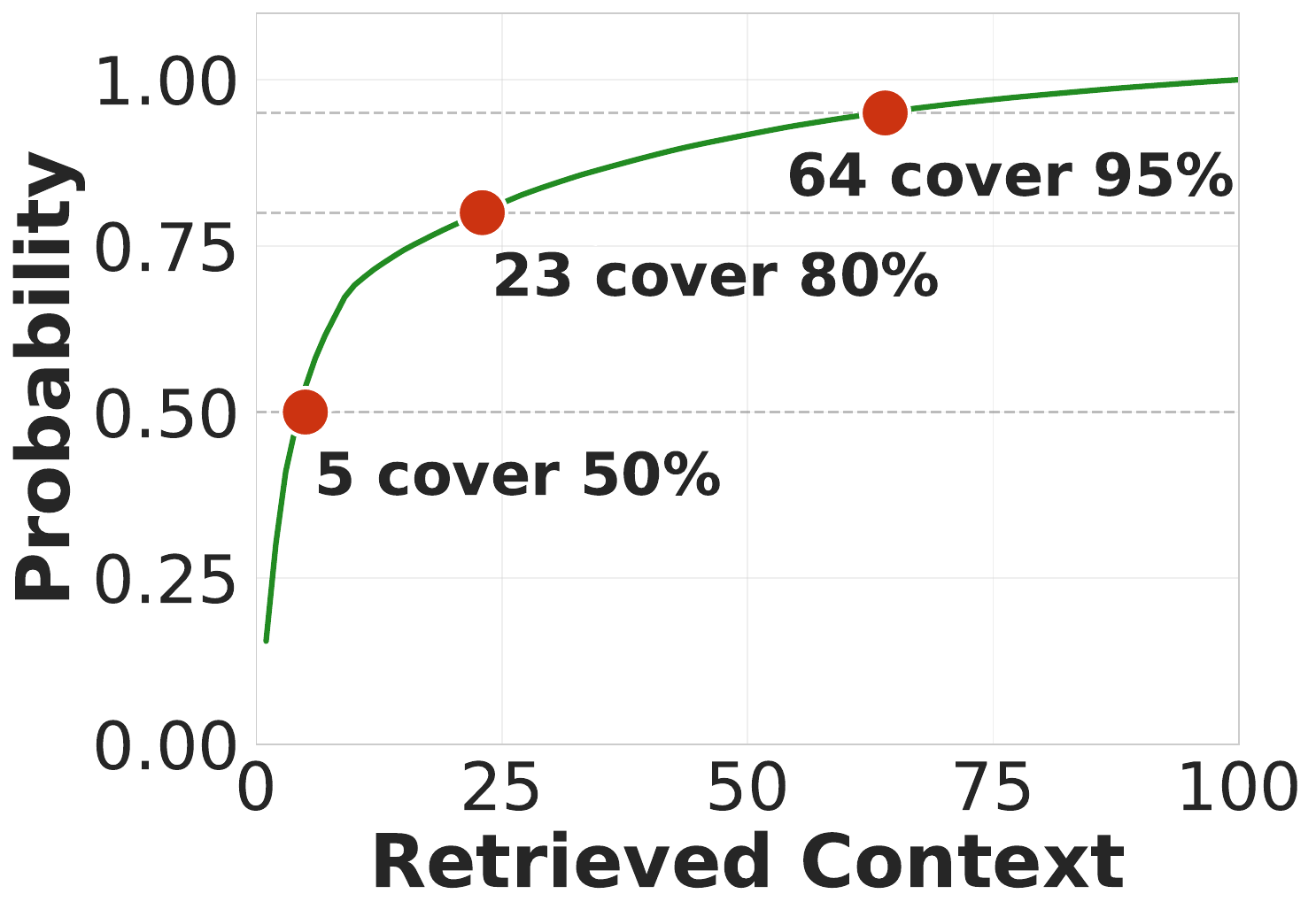}
        \caption{Low Skewness}
        \label{fig:cdf_cwq_low}
    \end{subfigure}
    \caption{\textbf{Cumulative Probability of Context Scores.} (a)(b) Results on the WebQSP Dataset (c)(d) Results on the CWQ Dataset.}
    \label{fig:cdf} 
\end{figure}

Building upon the previously observed variations in score skewness across queries, we visualize their cumulative probability on the WebQSP and CWQ dataset, as shown in Figure \ref{fig:cdf}. 
These plots reveal two primary types of cumulative behavior: high skewness and low skewness.
For the former, where scores are concentrated among just a few retrieved contexts, the cumulative probability rises steeply. Thus, a high cumulative probability is achieved with only a minimal number of contexts (e.g., Figure \ref{fig:cdf_webqsp_high} illustrates reaching 95\% cumulative probability with just 5 triples).
In contrast, the low skewness where scores are more evenly distributed among more contexts, exhibit a slower rise in cumulative probability, requiring more to reach the same value (e.g., Figure \ref{fig:cdf_webqsp_low}, 95\% requires 63). 
Similar behavior was also observed in the CWQ dataset.
This significant variation in the number of contexts required for a given cumulative probability directly reflects the differences in underlying distribution skewness of scores. This observation inspires our cumulative threshold-based method, which leverages the context number needed to achieve a preset cumulative probability to inform dynamic routing decisions.

\subsection{Routing Across Cumulative Probability}
\begin{figure}[t] 
    \centering 
    \begin{subfigure}[b]{0.49\linewidth} 
        \centering 
        \includegraphics[width=\linewidth]{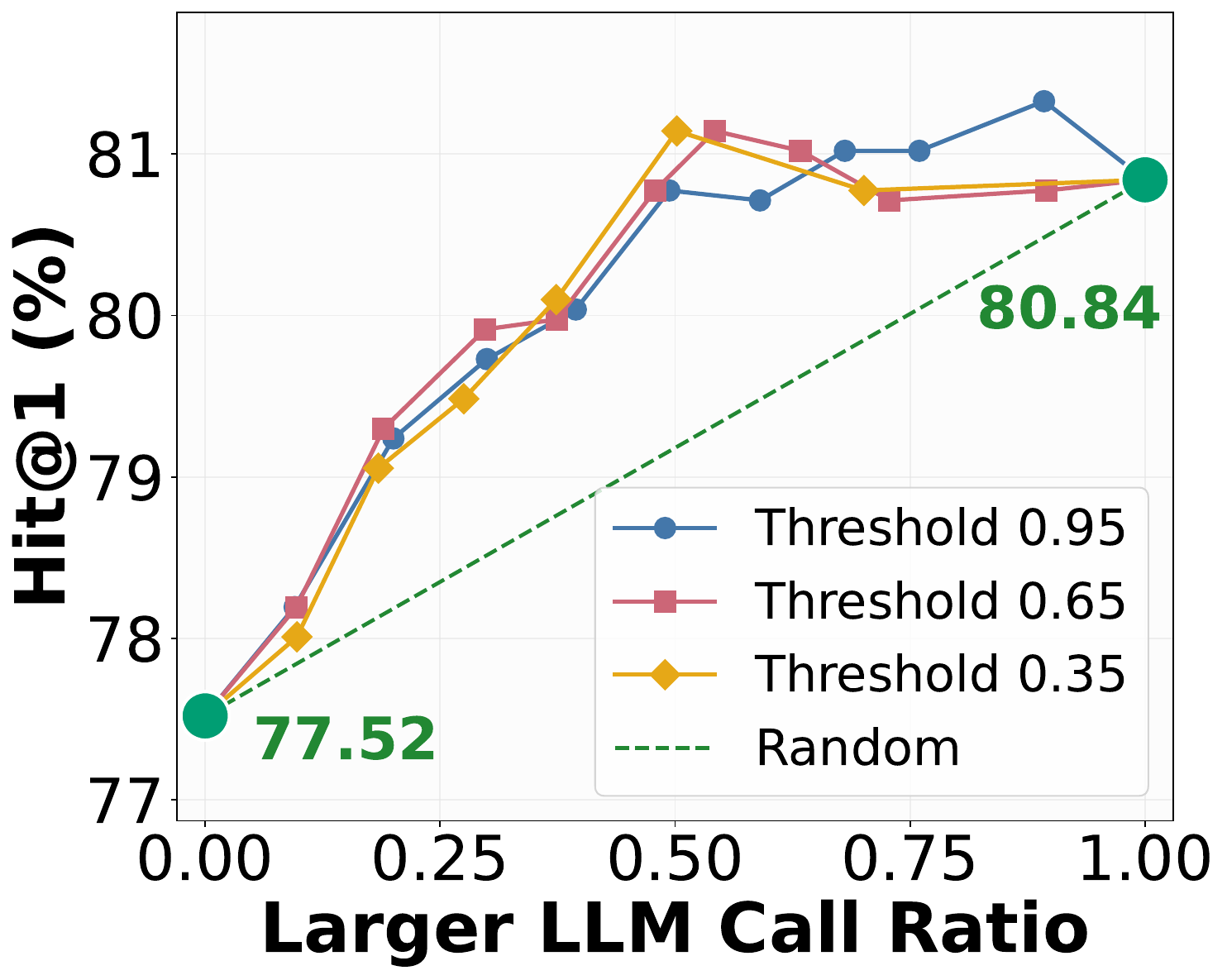}
        \caption{WebQSP Hit@1} 
        \label{fig:thres_results:webqsp:hit}
    \end{subfigure}
    \hfill 
    \begin{subfigure}[b]{0.49\linewidth}
        \centering
        \includegraphics[width=\linewidth]{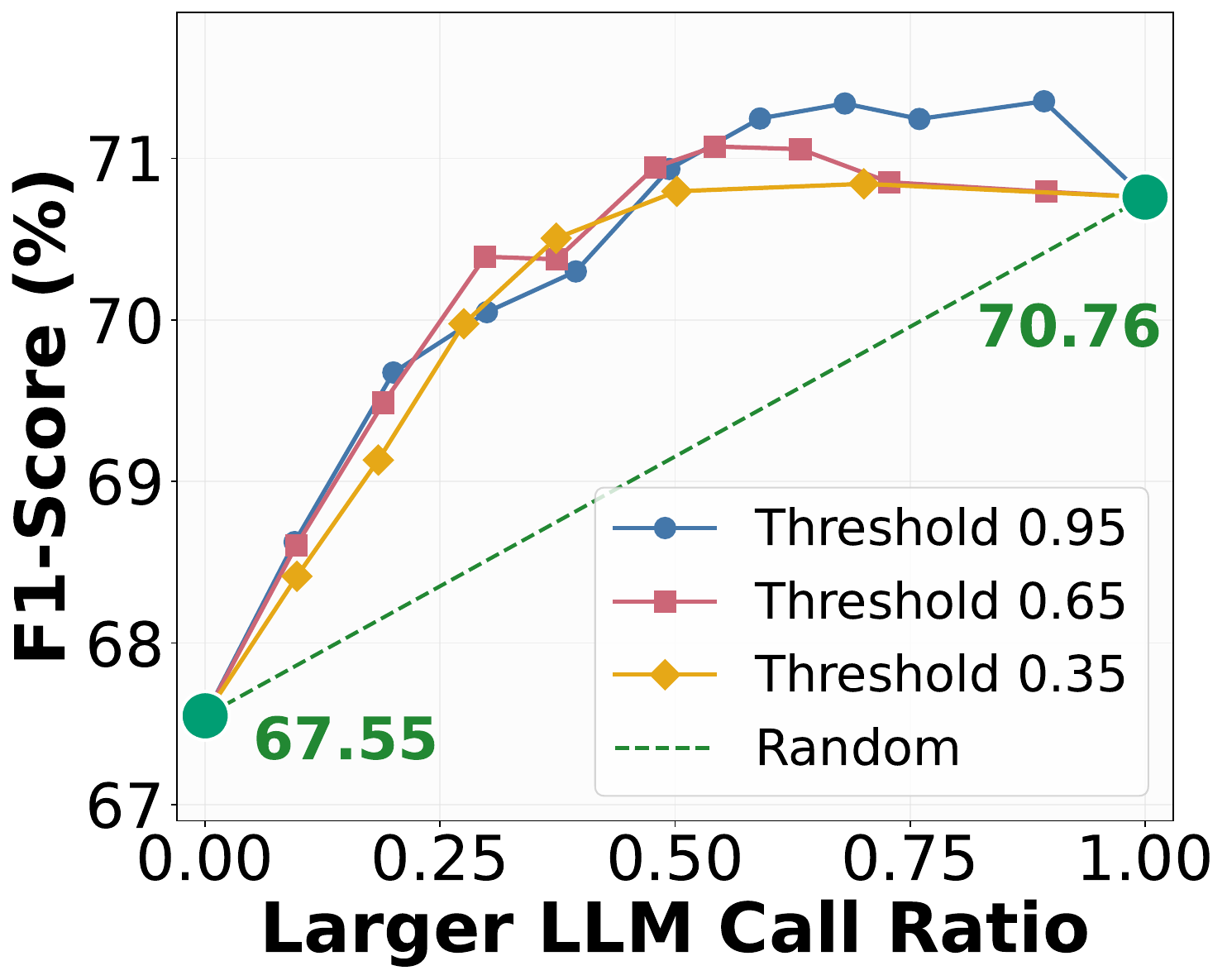}
        \caption{WebQSP F1-Score}
        \label{fig:thres_results:webqsp:f1}
    \end{subfigure}
    \begin{subfigure}[b]{0.49\linewidth}
        \centering
        \includegraphics[width=\linewidth]{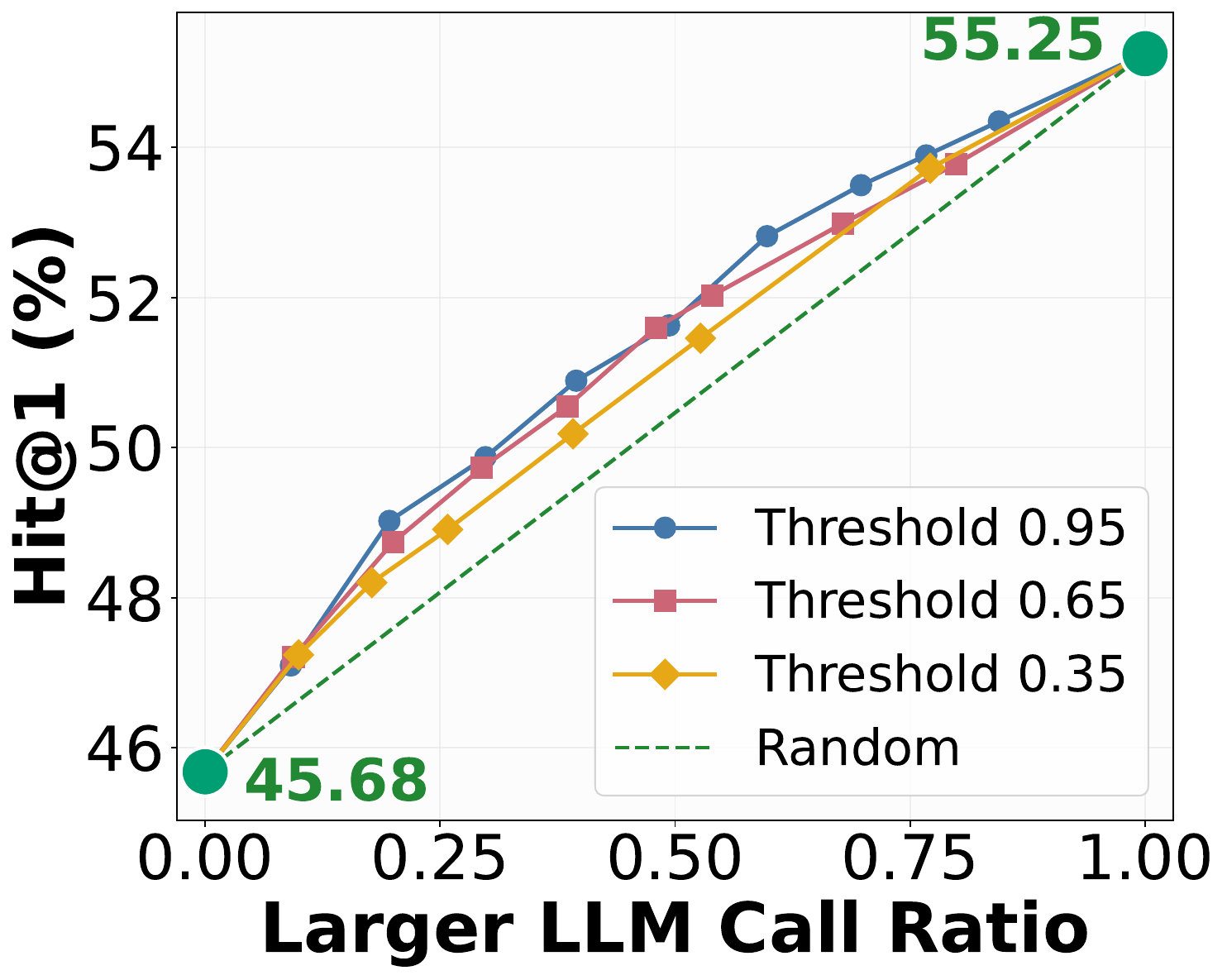}
        \caption{CWQ Hit@1}
        \label{fig:thres_results:cwq:hit}
    \end{subfigure}
    \hfill 
    \begin{subfigure}[b]{0.49\linewidth}
        \centering
        \includegraphics[width=\linewidth]{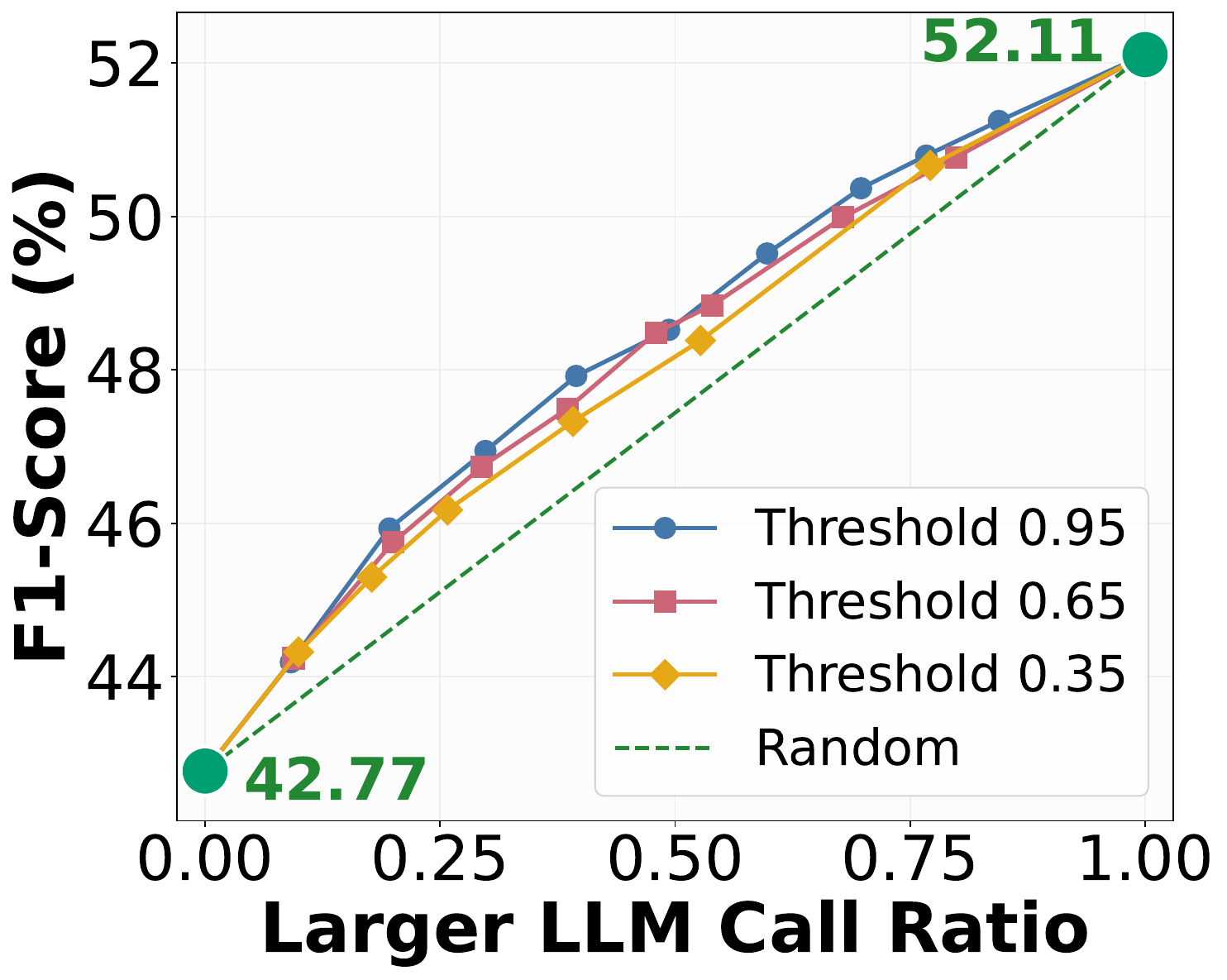}
        \caption{CWQ F1-Score}
        \label{fig:thres_results:cwq:f1}
    \end{subfigure}
    \caption{\textbf{Results of Cumulative Threshold-based Routing with Different $P$.} (a)(b) on WebQSP and (c)(d) on CWQ between Qwen7b and Qwen72b with Cumulative Probability $P$ in \{0.35, 0.65, 0.95\}.}
    \label{fig:thres_results} 
\end{figure}

We conduct an experiment to explore the impact of the probability $P$ in the cumulative threshold-based method, as shown in Figure \ref{fig:thres_results}. 
Overall, results show that our training-free methods consistently outperform the random mixing baseline for cumulative probability $P$ ranging from 0.35 to 0.95, demonstrating its robustness.
For example, on the WebQSP dataset, compared to using the LLM exclusively for inference, our methods reduce cost over 50\% without any loss in performance. 
Highly skewed distributions of contexts scores reach given $P$ with very few examples, while those with low skewness require many more.
The probability choice is crucial. 
A low probability $P$ is easily triggered by queries with highly skewed scores, but it also allows those with low skewness to pass, thereby diminishing the router's discriminative power. 
As the results illustrate, $P = 0.95$ steadily outperforms $P = 0.35$. 

\subsection{Correlation Analysis} 
\label{appendix:cor}

\begin{figure}[htbp] 
    \centering
    \begin{subfigure}[b]{0.49\linewidth} 
        \centering 
        \includegraphics[width=\linewidth]{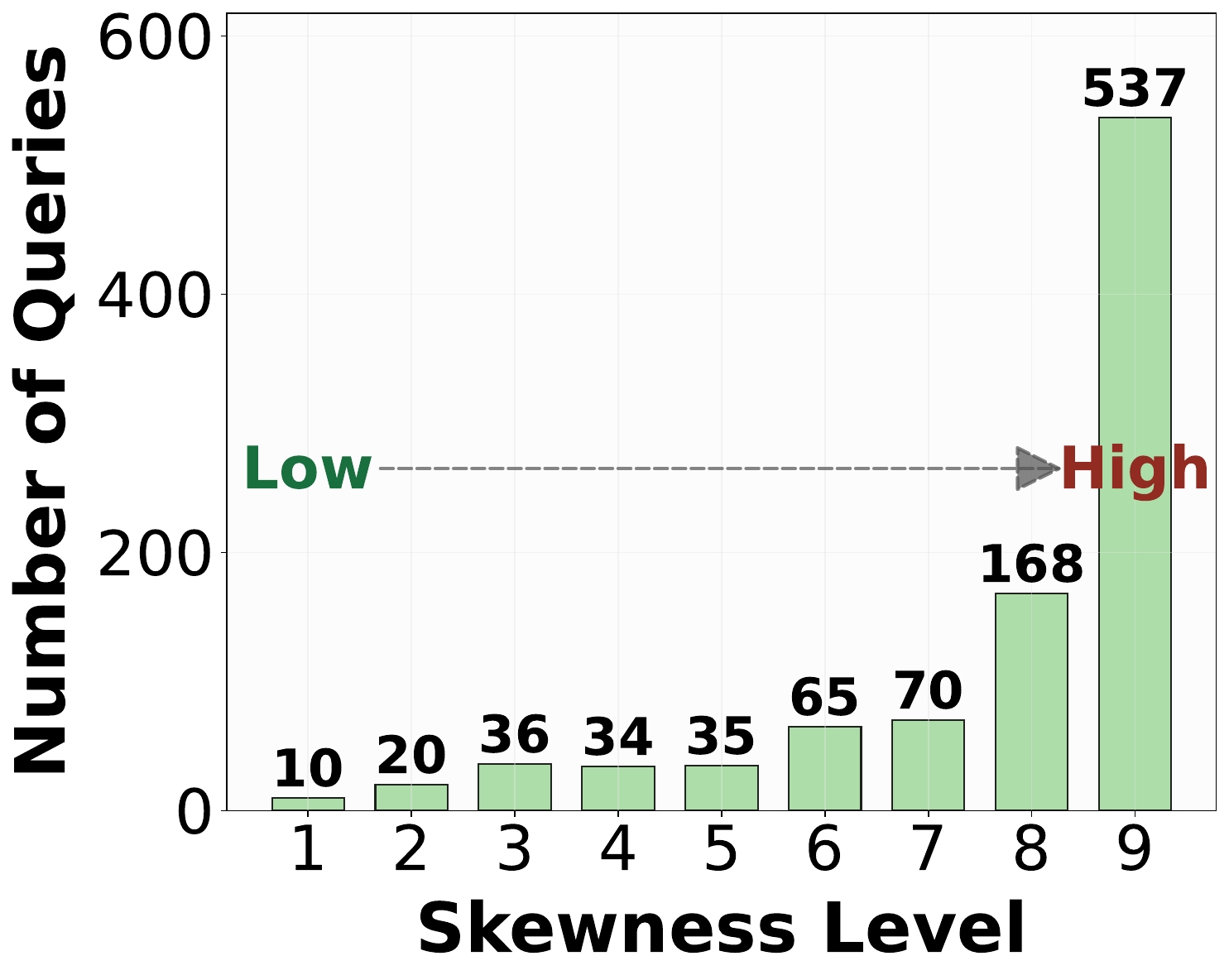}
        \caption{WebQSP Dataset}
        \label{fig:correlation:webqsp}
    \end{subfigure}
    \hfill 
    \begin{subfigure}[b]{0.49\linewidth} 
        \centering 
        \includegraphics[width=\linewidth]{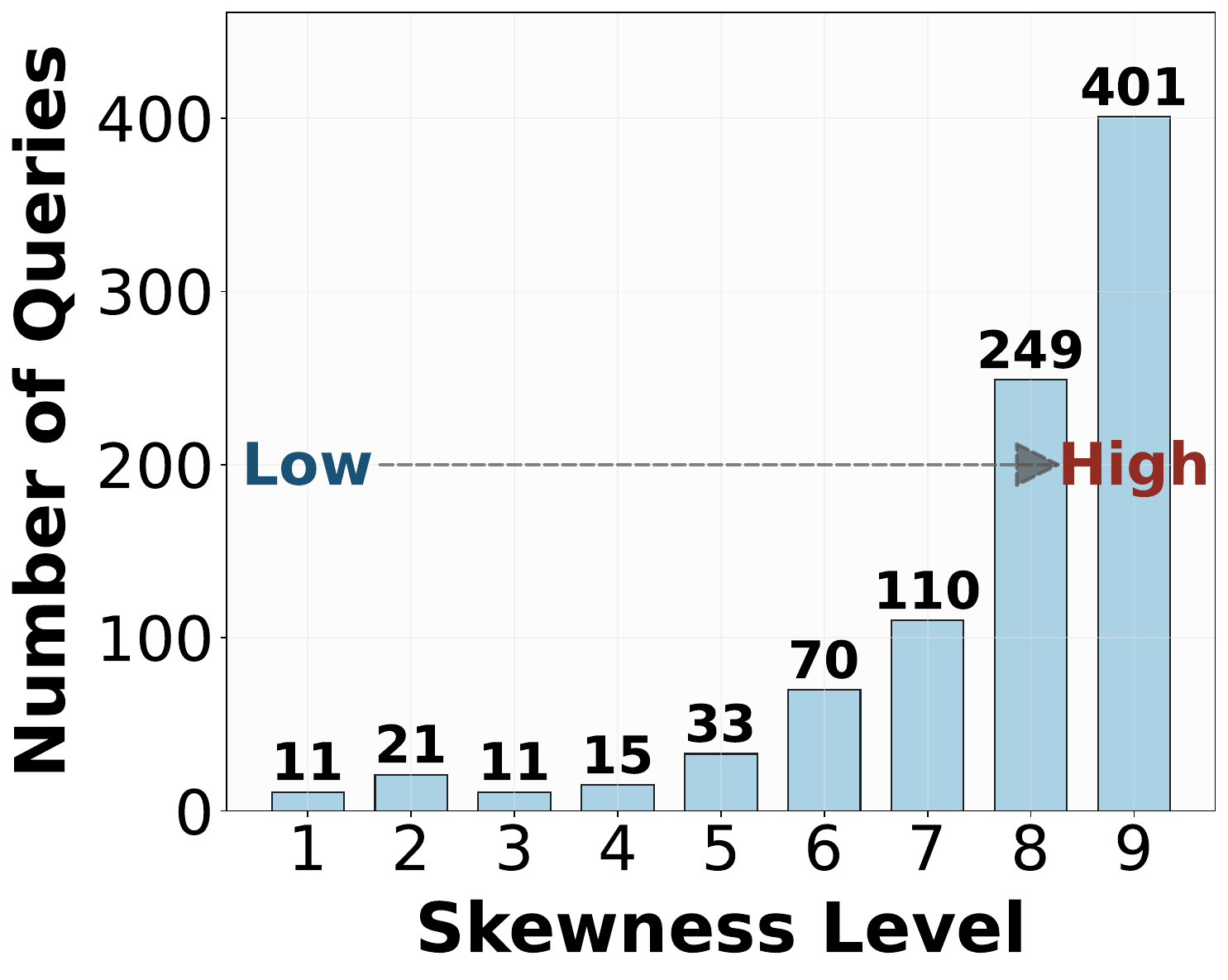}
        \caption{CWQ Dataset}
        \label{fig:correlation:cwq}
	\end{subfigure}
    \caption{\textbf{Simple Query Number Across Skewness.}}
    \label{fig:correlation} 
\end{figure}

In the absence of supervision, there are no explicit labels to directly assess query difficulty. Nevertheless, in the RAG scenario, the retrieval quality directly impacts the answer generation of LLMs.
The retrieval of clear and sufficient contexts inherently lowers the difficulty of answering the given query.
A query is trivially simple if it can be answered with a single triple that appears right next to it in the prompt.
Thus, to assess query difficulty, we use a straightforward metric: we check whether the top-ranked context in retrieved set contains the answer for given query. 
Simple query requires only a single knowledge context to answer. If the Top-$1$ contains real answer, we consider this query is simple.
In this case, the scorer has accurately identified the context critical for reasoning with high confidence. 

Next, we investigate the relationship between score skewness of retrieved contexts and query difficulty. 
To this end, we extract queries whose answers appear in the top-scoring context and compute the areas under their scores.
Smaller areas correspond to higher skewness. We partition the area from their minimum to maximum into several contiguous intervals, each corresponding to a distinct level of skewness.
Figure \ref{fig:correlation} illustrates the count of simple queries across different skewness levels.
Overall, a consistent trend was observed in both datasets, which queries are predominantly concentrated at higher skewness levels. 
This allows us to select simple queries based on their skewness. 
The finding further reveals the strong correlation between query difficulty and score skewness, which can serve as effective indicator for routing LLMs to balance performance and cost. 

\begin{figure}[t]
    \centering 
    \begin{subfigure}[b]{0.49\linewidth} 
        \centering 
        \includegraphics[width=\linewidth]{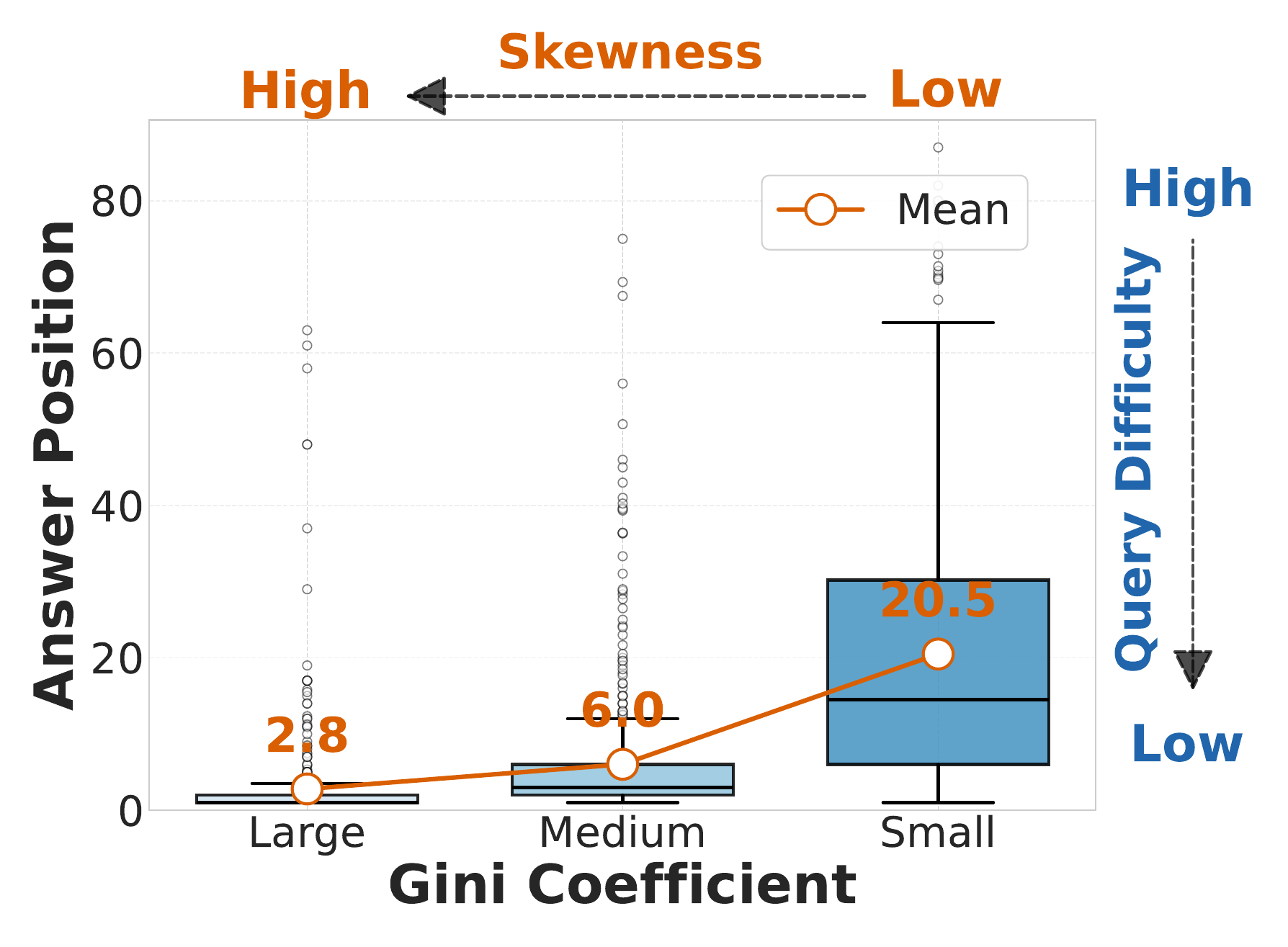}
        \caption{WebQSP Gini-based} 
        \label{fig:webqsp_gini} 
    \end{subfigure}
    \hfill 
    \begin{subfigure}[b]{0.49\linewidth}
        \centering
        \includegraphics[width=\linewidth]{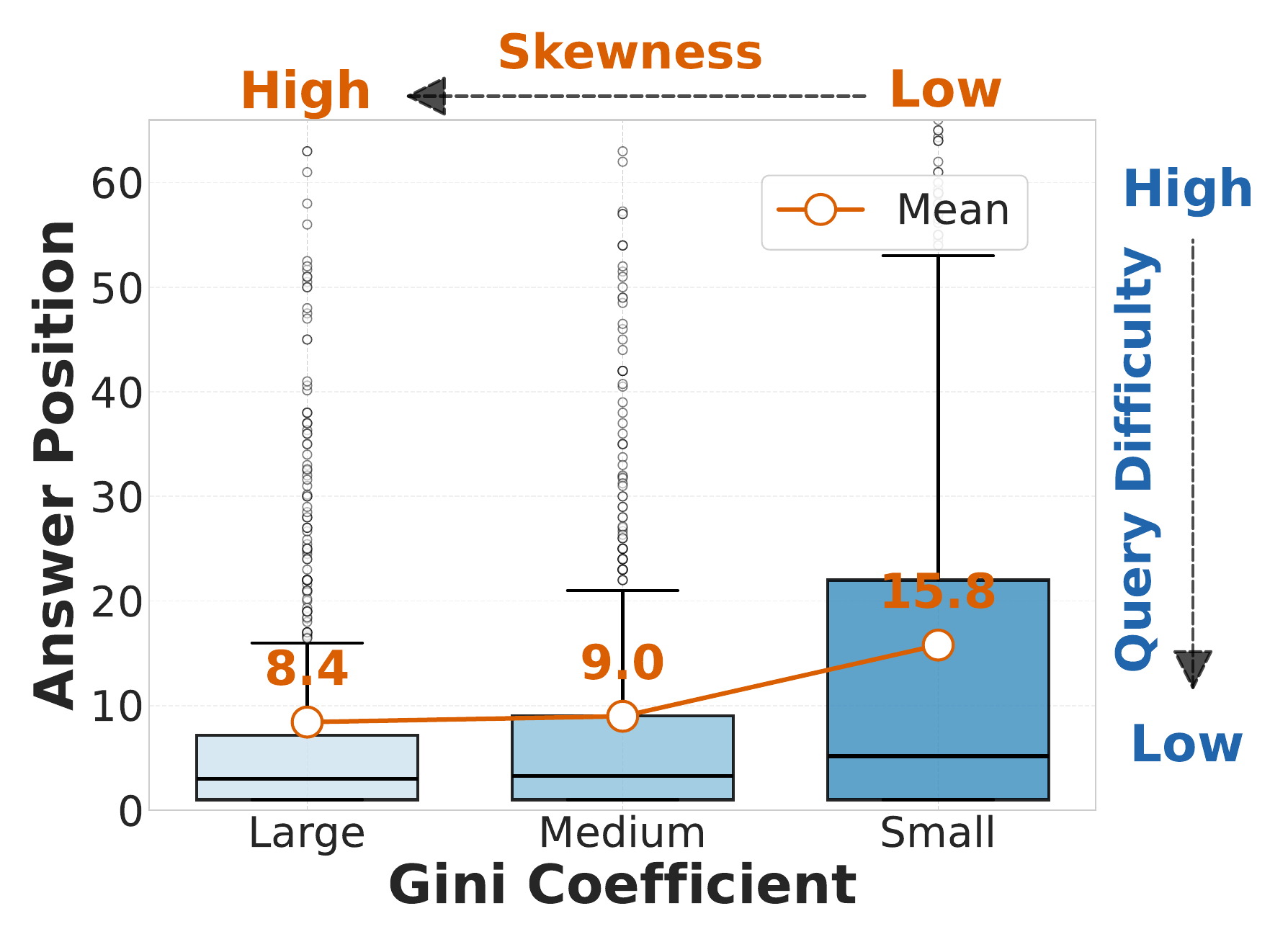}
        \caption{CWQ Gini-based}    
        \label{fig:cwq_gini}
    \end{subfigure}
    \begin{subfigure}[b]{0.49\linewidth}
        \centering
        \includegraphics[width=\linewidth]{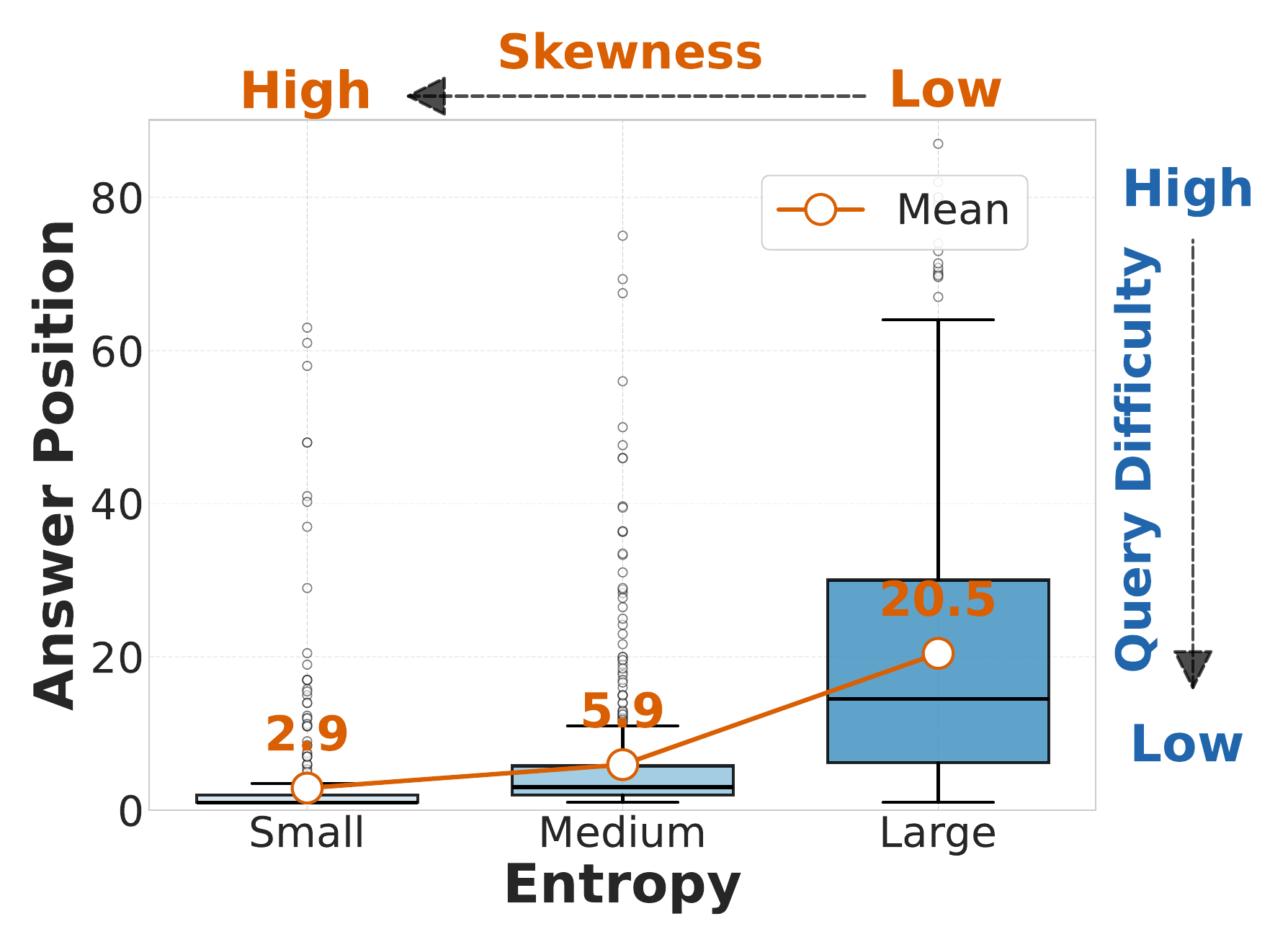}
        \caption{WebQSP Entropy-based}
        \label{fig:webqsp_entropy}
    \end{subfigure}
    \hfill 
    \begin{subfigure}[b]{0.49\linewidth}
        \centering
        \includegraphics[width=\linewidth]{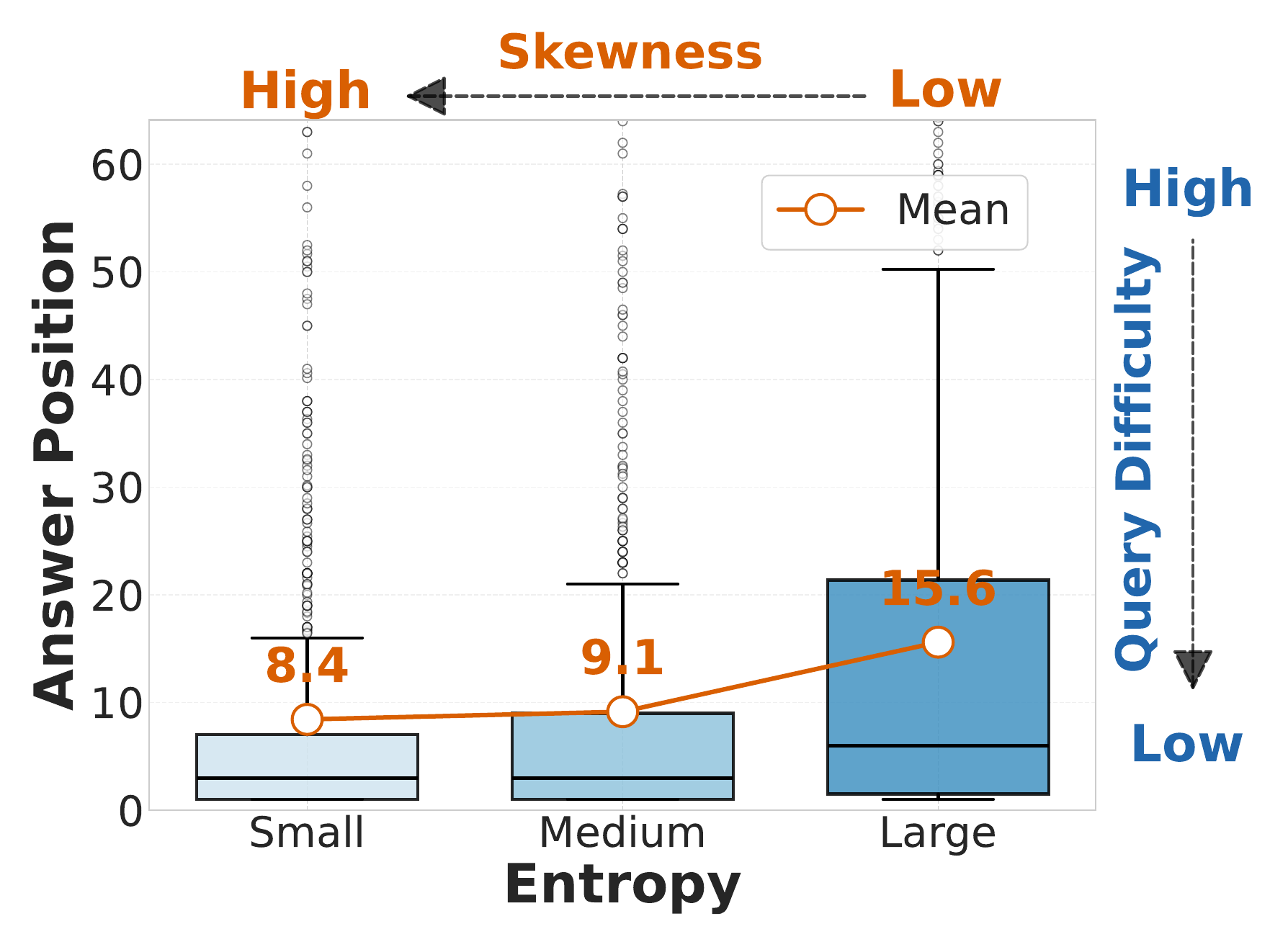}
        \caption{CWQ Entropy-based}
        \label{fig:cwq_entropy}
    \end{subfigure}
     \begin{subfigure}[b]{0.49\linewidth}
        \centering
        \includegraphics[width=\linewidth]{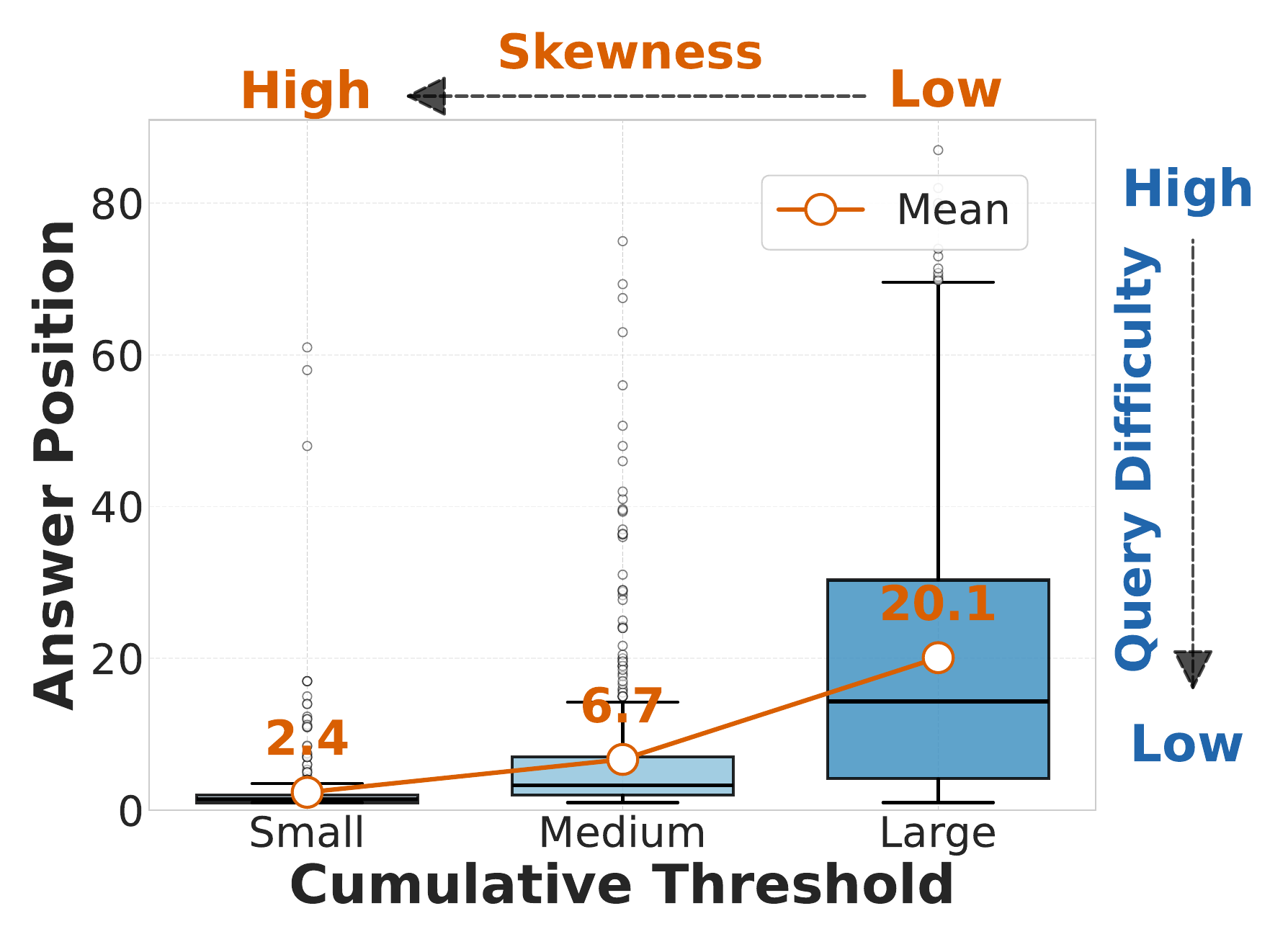}
        \caption{WebQSP Cumu-based}
        \label{fig:webqsp_thres}
    \end{subfigure}
    \hfill 
    \begin{subfigure}[b]{0.49\linewidth}
        \centering
        \includegraphics[width=\linewidth]{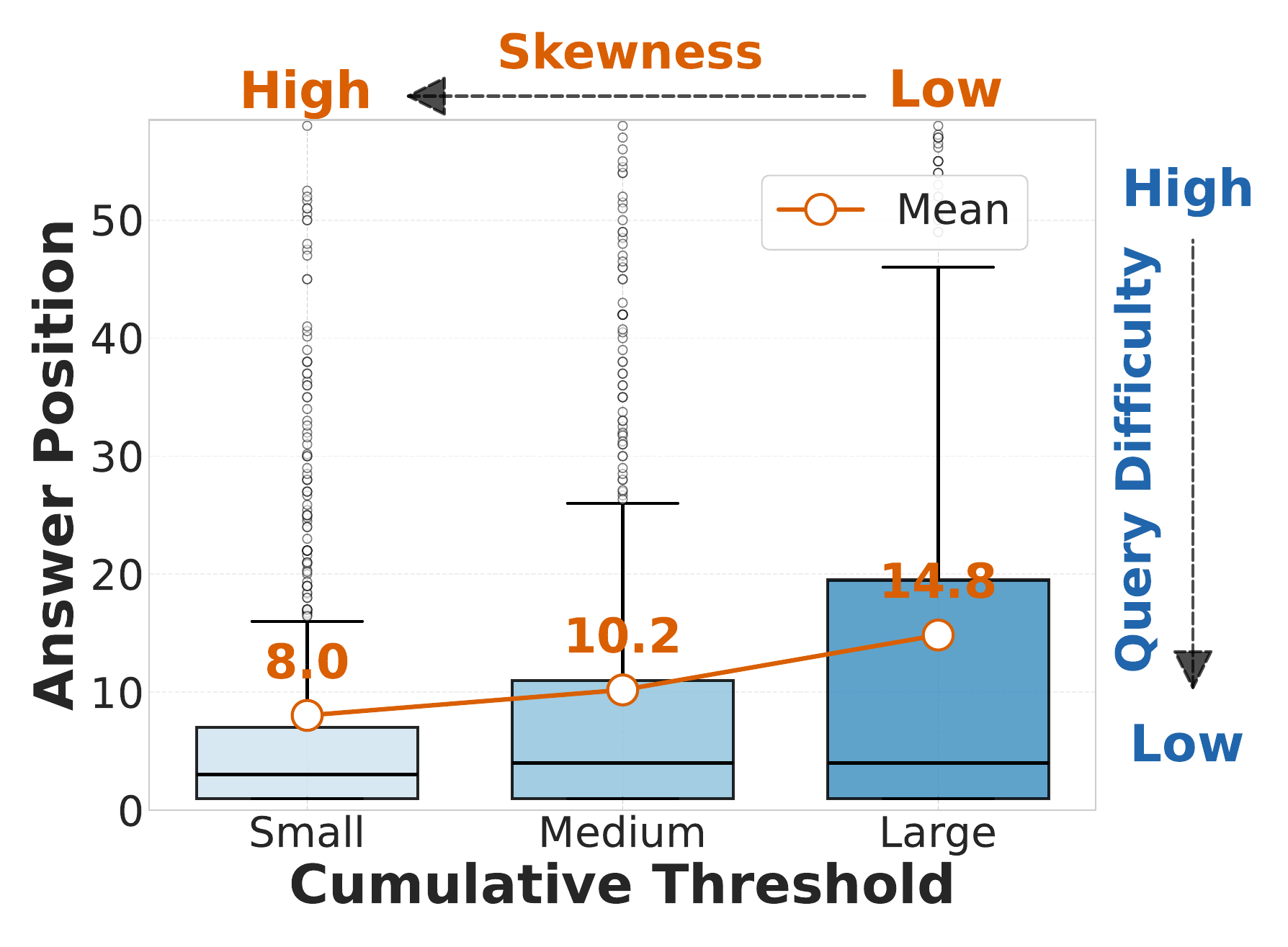}
        \caption{CWQ Cumu-based}
        \label{fig:cwq_thres}
    \end{subfigure}
    \caption{\textbf{Correlation Between Query Difficulty and Score Skewness.} The subfigures show results for different skewness calculation metrics on WebQSP and CWQ dataset: (a)(b) Gini Coefficient, (b)(d) Entropy, (e)(f) Cumulative Threshold. The red line shows the mean answer position of each group.}
    \label{fig:appendix_correlation} 
\end{figure}
Furthermore, our gini-based, entropy-based and cumulative-based metrics for distribution skewness exhibit a trend consistent with the area-based method, as depicted in Figure \ref{fig:appendix_correlation}. 
The rank of the answer-containing context can indicate query difficulty, where a lower position typically corresponds to a more complex, multi-hop reasoning scenario.
We can observe a strong correlation where as the score distribution becomes more skewed, the rank of answer-containing context is higher, consequently, lower query difficulty. 
This reinforces the core insight: score skewness is a reliable indicator of query difficulty, enabling a dynamic routing strategy to balance inference performance and cost across different LLMs.

\begin{table*}[t]
\centering
\small
\setlength{\tabcolsep}{3.1pt}
\begin{tabular}{llccccccc}
\toprule
Dataset & Method                   & 0\%    & 20\%                      & 40\%                      & 60\%                      & 80\%                      & 100\%   & Avg. Eff. \\
\midrule
\multirow{7}{*}{WebQSP}
  & Random Routing          & 77.52  & 78.18                     & 78.85                     & 79.51                     & 80.18                     & 80.84   & - \\
  & RouteLLM    [ICLR, 25]  & 77.52  & 78.56 (+0.38)             & 79.18 (+0.33)             & 79.55 (+0.04)             & 79.98 (-0.20)             & 80.84   & +0.14 \\
  & GraphRouter [ICLR, 25]  & 77.52  & 78.26 (+0.08)             & 79.30 (+0.45)             & 80.22 (+0.71)             & 80.71 (+0.53)             & 80.84   & +0.44 \\
  & Ours (Area-based)       & 77.52  & 78.26 (+0.08)             & \underline{80.10 (+1.25)} & \underline{81.08 (+1.57)} & \textbf{81.14 (+0.96)}    & 80.84   & +0.97 \\
  & Ours (Gini-based)       & 77.52  & \textbf{79.48 (+1.30)}    & 79.98 (+1.13)             & \textbf{81.20 (+1.69)}    & \underline{80.96 (+0.78)} & 80.84   & \textbf{+1.23} \\
  & Ours (Entropy-based)    & 77.52  & \textit{79.18 (+1.00)}    & \textbf{80.34 (+1.49)}    & \underline{81.08 (+1.57)} & \textit{80.77 (+0.59)}    & 80.84   & \textit{+1.16} \\
  & Ours (Cumulative-based) & 77.52  & \underline{79.24 (+1.06)} & \textit{80.04 (+1.19)}    & 80.71 (+1.20)             & 80.71 (+0.53)             & 80.84   & \underline{+1.20} \\
\midrule
\multirow{7}{*}{CWQ}
  & Random Routing          & 45.68  & 47.59                     & 49.51                     & 51.42                     & 53.34                     & 55.25   & - \\
  & RouteLLM    [ICLR, 25]  & 45.68  & 47.81 (+0.22)             & 50.41 (+0.90)             & 52.00 (+0.58)             & 53.61 (+0.27)             & 55.25   & +0.49 \\
  & GraphRouter [ICLR, 25]  & 45.68  & 47.83 (+0.24)             & 50.10 (+0.59)             & 52.02 (+0.60)             & 53.78 (+0.44)             & 55.25   & +0.47 \\
  & Ours (Area-based)       & 45.68  & 48.06 (+0.47)             & 49.59 (+0.08)             & 52.02 (+0.60)             & 53.95 (+0.61)             & 55.25   & +0.44 \\
  & Ours (Gini-based)       & 45.68  & \underline{48.94 (+1.35)} & \textbf{50.92 (+1.41)}    & \textit{52.53 (+1.11)}    & \textbf{54.15 (+0.81)}    & 55.25   & \underline{+1.17} \\
  & Ours (Entropy-based)    & 45.68  & \textit{48.74 (+1.15)}    & \textit{50.72 (+1.21)}    & \underline{52.70 (+1.28)} & \underline{54.01 (+0.67)} & 55.25   & \textit{+1.08} \\
  & Ours (Cumulative-based) & 45.68  & \textbf{49.02 (+1.43)}    & \underline{50.89 (+1.38)} & \textbf{52.82 (+1.40)}    & \textit{53.89 (+0.55)}    & 55.25   & \textbf{+1.19} \\
\bottomrule
\end{tabular}
\caption{\textbf{Routing Between Qwen2.5 Models: 7B as the Small LLM and 72B as the Large LLM with Area-based Routing Method.} The best scores are highlighted with \textbf{bold}, the second-best scores are highlighted with \underline{underline} and the third-best scores are indicated in \textit{italics}.}
\label{tab:main_qwen_area}
\end{table*}

\begin{table*}[h]
\centering
\small
\setlength{\tabcolsep}{3.1pt}
\begin{tabular}{llccccccc}
\toprule
Dataset & Method                   & 0\%     & 20\%                      & 40\%                      & 60\%                      & 80\%                      & 100\%    & Avg. Eff. \\
\midrule
\multirow{7}{*}{WebQSP}
  & Random Routing          & 78.56   & 79.68                     & 80.80                     & 81.91                     & 83.03                     & 84.15    & - \\
  & RouteLLM    [ICLR, 25]  & 78.56   & 79.73 (+0.05)             & 80.47 (-0.33)             & 81.82 (-0.09)             & 82.86 (-0.17)             & 84.15    & -0.14 \\
  & GraphRouter [ICLR, 25]  & 78.56   & 79.67 (-0.01)             & 80.65 (-0.15)             & 81.39 (-0.52)             & 83.35 (+0.32)             & 84.15    & -0.09 \\
  & Ours (Area-based)       & 77.52   & 80.22 (+0.54)             & \underline{81.76 (+0.96)} & 82.43 (+0.52)             & \textbf{83.91 (+0.88)}    & 80.84    & +0.73 \\
  & Ours (Gini-based)       & 78.56   & \underline{81.33 (+1.65)} & \textit{81.57 (+0.77)}    & \textbf{82.62 (+0.71)}    & 83.35 (+0.32)             & 84.15    & \textit{+0.86} \\
  & Ours (Entropy-based)    & 78.56   & \textit{81.08 (+1.40)}    & \textbf{82.00 (+1.20)}    & \underline{82.49 (+0.58)} & \underline{83.66 (+0.63)} & 84.15    & \textbf{+0.95} \\
  & Ours (Cumulative-based) & 78.56   & \textbf{81.57 (+1.89)}    & 81.27 (+0.47)             & \textbf{82.62 (+0.71)}    & \textit{83.60 (+0.57)}    & 84.15    & \underline{+0.91} \\
\midrule
\multirow{7}{*}{CWQ}
  & Random Routing          & 49.90   & 51.51                     & 53.12                     & 54.72                     & 56.33                     & 57.94    & - \\
  & RouteLLM    [ICLR, 25]  & 49.90   & 51.26 (-0.25)             & 53.61 (+0.49)             & 55.03 (+0.31)             & 56.75 (+0.42)             & 57.94    & +0.24 \\
  & GraphRouter [ICLR, 25]  & 49.90   & 50.84 (-0.67)             & 52.79 (-0.33)             & 54.80 (+0.08)             & 56.16 (-0.17)             & 57.94    & -0.27 \\
  & Ours (Area-based)       & 49.90   & 52.25 (+0.74)             & 53.87 (+0.75)             & 55.45 (+0.73)             & \textit{56.73 (+0.40)}    & 57.94    & +0.66 \\
  & Ours (Gini-based)       & 49.90   & \underline{52.65 (+1.14)} & \textbf{55.00 (+1.88)}    & \underline{56.16 (+1.44)} & \textbf{57.04 (+0.71)}    & 57.94    & \textbf{+1.29} \\
  & Ours (Entropy-based)    & 49.90   & \textit{52.51 (+1.00)}    & \underline{54.89 (+1.77)} & \textit{56.07 (+1.35)}    & \underline{56.78 (+0.45)} & 57.94    & \textit{+1.14} \\
  & Ours (Cumulative-based) & 49.90   & \textbf{52.68 (+1.17)}    & \textit{54.77 (+1.65)}    & \textbf{56.41 (+1.69)}    & 56.61 (+0.28)             & 57.94    & \underline{+1.20} \\
\bottomrule
\end{tabular}
\caption{\textbf{Routing Between Llama3.1 Models: 8B as the Small LLM and 70B as the Large LLM with Area-based Routing Method.} The best scores are highlighted with \textbf{bold} and the second-best scores are highlighted with \underline{underline} and the third-best scores are indicated in \textit{italics}.}
\label{tab:main_llama_area}
\end{table*}

\subsection{Results of Area-based Routing Method}
To evaluate the area-based routing method, we conduct experiments on the WebQSP and CWQ datasets using Qwen2.5-7B-Instruct and Qwen2.5-72B-Instruct.
The area under the context score curve naturally reflects skewness. High skewness corresponds to a rapid decline, yielding a smaller area enclosed with the coordinate axes, whereas low skewness indicates gradual declines with more high-scoring contexts, thus forming a larger area. 
Despite being simple and intuitive, this area-based method proves effective, as demonstrated by Table \ref{tab:main_qwen_area}.
Overall, the area-based method consistently outperforms all baselines on the WebQSP dataset.
Specifically, although slightly less competitive than other three methods, it still surpasses RouteLLM nearly 7x and GraphRouter over 2x.
On the CWQ dataset, the area-based method performs on par with RouteLLM and GraphRouter, while requiring no training.
These findings underscore the potential of area-based method as an effective and training-free solution for query routing in knowledge-intensive tasks.

Table~\ref{tab:main_llama_area} presents the routing results between Llama3.1-8B-Instruct and Llama3.1-70B-Instruct.
The area-based method demonstrates strong plug-and-play generalization, requiring no training while still delivering consistent gains over baseline routers.
On the WebQSP dataset, the area-based method achieves an Average Effectiveness of 0.73, even when RouteLLM and GraphRouter fail.
On the CWQ dataset, the area-based method remains effective, outperforming the strong baseline RouteLLM by nearly 3x.
These results highlight the area-based method as a training-free, plug-and-play and readily deployable routing strategy with strong generalization across both datasets and model families.

\onecolumn
\subsection{Prompts}
\label{appendix:prompt}
The following is the detailed prompt template used in SubgraphRAG for all experiments.
Due to the effects of positional encoding in the self-attention mechanism that tokens appearing later in the sequence tend to receive higher attention weights, we place triples in ascending order. 
\begin{tcolorbox}[title=SubgraphRAG Prompt]
	\noindent\textbf{System:} 
    
    Based on the triplets retrieved from a knowledge graph, please answer the question. Please return formatted answers as a list, each prefixed with "ans:".

    \noindent\rule{\textwidth}{0.7pt} 

    \noindent\textbf{User:} 

    \noindent\textit{Triplets:}\\
    (m.011zsc4\_, organization.leadership.organization, San Francisco Giants)\\
    (m.0crtd80, sports.sports\_league\_participation.league, National League West)\\
    \ldots\\
    (San Francisco Giants, time.participant.event, 2014 Major League Baseball season)\\
    (San Francisco Giants, time.participant.event, 2012 Major League Baseball season)\\
    (AT\&T Park, location.location.events, 2010 World Series)\\
    (San Francisco Giants, sports.professional\_sports\_team.owner\_s, Bill Neukom)\\
    (San Francisco Giants, time.participant.event, 2010 Major League Baseball season)\\
    (San Francisco Giants, sports.sports\_team.championships, 2010 World Series)\\
    (San Francisco Giants, time.participant.event, 2012 World Series)\\
    (Crazy Crab, sports.mascot.team, San Francisco Giants)\\
    (San Francisco Giants, time.participant.event, 2010 World Series)\\
    (San Francisco Giants, sports.sports\_team.championships, 2012 World Series)\\
    (San Francisco Giants, sports.sports\_team.team\_mascot, Crazy Crab)\\
    (San Francisco Giants, sports.sports\_team.championships, 2014 World Series)\\
    (Lou Seal, sports.mascot.team, San Francisco Giants)\\

    \noindent\textit{Question:}
    What year did the team with mascot named Lou Seal win the World Series?

    \noindent\rule{\textwidth}{0.7pt} 

    \noindent\textbf{Assistant:} 

    To find the year the team with mascot named Lou Seal won the World Series, we need to find the team with mascot named Lou Seal and then find the year they won the World Series. From the triplets, we can see that Lou Seal is the mascot of the San Francisco Giants. Now, we need to find the year the San Francisco Giants won the World Series. From the triplets, we can see that San Francisco Giants won the 2010 World Series and 2012 World Series and 2014 World Series. So, the team with mascot named Lou Seal (San Francisco Giants) won the World Series in 2010, 2012, and 2014. Therefore, the formatted answers are: 
    \\ans: 2014 World Series
    \\ans: 2012 World Series
    \\ans: 2010 World Series

    \noindent\rule{\textwidth}{0.7pt} 

    \noindent\textbf{User:} 

    Triplets:\\
    Question:\\
\end{tcolorbox}

\end{document}